\documentclass[pdftex,twocolumn,epjc3]{svjour3}          % twocolumn

\RequirePackage[T1]{fontenc}

\smartqed  % flush right qed marks, e.g. at end of proof
\usepackage{amsmath,amssymb,amsfonts,multicol}
\RequirePackage{graphicx}
\RequirePackage{mathptmx}      % use Times fonts if available on your TeX system
\RequirePackage{flushend}
\RequirePackage[numbers,sort&compress]{natbib}
\RequirePackage[colorlinks,citecolor=blue,urlcolor=blue,linkcolor=blue]{hyperref}

\begin{document}

\title{Gravitational waveforms from inspiral compact binaries in Hybrid metric-Palatini gravity}

%\subtitle{Do you have a subtitle?\\ If so, write it here}

\author{Polina Dyadina\thanksref{e1,addr1}
         %etc.
}

\thankstext{e1}{e-mail: guldur.anwo@gmail.com}

\institute{Sternberg Astronomical Institute, Lomonosov Moscow State University, Universitetsky Prospekt, 13, Moscow, 119234, Russia\label{addr1}
        }

\date{Received:  / Accepted: }
% The correct dates will be entered by the editor

\maketitle

\begin{abstract} 
In this study, gravitational waveforms emitted by inspiralling compact binary systems on quasicircular orbits in hybrid metric-Palatini gravity are computed in the lowest post-Newtonian approximation. By applying the stationary phase approximation, Fourier transforms of the tensor polarization modes are obtained, and correction terms in the amplitude and phase of gravitational waves relative to General Relativity results are derived. Moreover, post-Einsteinian parameters are identified, and potential constraints on the background value of the scalar field are obtained based on possible observations of the gravitational waves by the future ground-based gravitational wave detectors. Additionally, constraints on the background value of the scalar field are derived using updated observational data from the PSR J0737-3039 system. These latter constraints are comparable in order of magnitude to the best currently existing constraints, which were derived from observational data within the Solar System.

\end{abstract}

\section{Introduction}

The discovery of gravitational waves (GWs) has opened the possibility for probing physics in strong gravitational regimes and has marked the beginning of a new era in gravitational astronomy \cite{Abbott_2016}. The detection of the merger of two neutron stars was significant, confirming that the speed of gravity is close to the speed of light \cite{Abbott_2017}. With the successes of the LIGO and Virgo collaborations in observing GWs from binary systems, there are new opportunities to test Einstein's general relativity (GR) under dynamic conditions of strong gravitational fields. Despite GR's successes, it faces theoretical and observational limitations, prompting the development of extensions to the theory. There are a number of problems that are not fully explained in the framework of GR, for example, the accelerated expansion of the Universe \cite{Perlmutter1999, Riess1998} or phenomena manifesting themselves as hidden mass \cite{zwicky, oort}, the period of inflation \cite{Starobinsky1980, Guth1981, Linde1982}, and the impossibility of constructing a quantum field model of gravity \cite{Burgess_2004, Latosh_2020}. To solve these problems, modified theories of gravity are used.  Any theory of gravity should be verifiable. Most tests of gravitational theories are based on experimental data obtained in weak gravitational fields \cite{Will_2014, kopeikin, will1993theory}. Observations of GWs provide clean information about strong gravitational fields and high energy scales, making them an ideal tool for experimental testing of gravitational theories.

One of the simplest and most widely used methods for extending GR is f(R)-gravity \cite{Bergmann1968, Felice2010, Nojiri2017, Nojiri2011}. The f(R)-gravity approach modifies the Einstein-Hilbert action by replacing the Ricci scalar with an arbitrary function of curvature, offering a compelling way to explain both the inflationary period and the current accelerated expansion of the universe. The f(R)-theories can be divided into two classes: metric and Palatini ones. In the metric approach, the metric is the sole variable, while the Palatini approach also incorporates an independent affine connection as a variable. Although the metric f(R)-models effectively explain the accelerated expansion of the Universe, they have problems with description of solar system dynamics \cite{Chiba2003, Olmo2005, Olmo2007}. Nevertheless, there are a number of viable models that can overcome these difficulties \cite{odintsov1, odintsov2, odintsov3}. On the other hand, the Palatini approach also has limitations in accordance with observational data \cite{koivisto2006, koivisto206}.

To address these issues, hybrid metric-Palatini gravity (HMPG) was developed, combining the advantages  of both metric and Palatini approaches while avoiding their respective shortcomings \cite{Harko2012, Capozziello2015, Harko2020}. HMPG encompasses both the metric components (via the Einstein-Hilbert action) and the Palatini elements (through an arbitrary function of Palatini curvature). This model successfully explains both cosmological and solar system dynamics without requiring additional screening mechanisms. Moreover, it admits a scalar-tensor representation, which simplifies its theoretical treatment.

HMPG has been extensively explored in various studies. For a comprehensive overview of this research, see \cite{Harko2020}. In this concise introduction, we summarize the primary research directions pursued within this framework. It's important to emphasize that HMPG has been analyzed across various scales and gravitational regimes. In cosmological contexts, the theory shows strong consistency with observational data \cite{Boehmer2013, Lima2016, Capozziello2013}. When applied to galactic scales, HMPG effectively models rotation curves with significantly reduced reliance on dark matter \cite{Capozziello2013a} and also addresses the virial mass discrepancies observed in galaxy clusters \cite{Capozziello13}. Furthermore, the theory is in agreement with observations within the solar system \cite{Leanizbarrutia2017}. The applicability of HMPG in the weak-field regime has been tested using the parameterized post-Newtonian formalism \cite{Dyadina2019, Dyadina_2022}. The reliability of the theory was further confirmed by its application to stronger gravitational fields, such as in binary pulsars \cite{Dyadina2018, Avdeev2020}. Additionally, the physical characteristics of neutron, Bose-Einstein condensate, and quark stars were explored in HMPG \cite{Danil2017}. In the strong-field limit, static spherically symmetric black hole solutions were obtained numerically \cite{Danila2019}, and the possibility of deriving stable spherically symmetric analytical solutions within HMPG has been discussed \cite{Bronnikov2020}. Moreover, accretion onto a static spherically symmetric black hole in the HMPG was investigated \cite{Dyadina_2024}. In addition, in a number of works, a generalized version of the HMPG was considered \cite{Rosa2017, Rosa20, Rosa2021}.

Gravitational waveforms have previously been obtained in a number of modified theories of gravity, including the massive Brans-Dicke theory \cite{Liu_2020}, Horndeski theory \cite{Higashino_2023}, metric \(f(R)\) models, and screened modified gravity \cite{Liu_2018}. HMPG has been partially explored in the context of gravitational waves. In particular, the number of polarization modes has been determined. It was found that HMPG predicts four polarizations: the tensor plus mode \(h_+\), the tensor cross mode \(h_\times\), the scalar breathing mode \(h_b\), and the scalar longitudinal mode \(h_L\) \cite{kausar2018}. Moreover, it was shown that the number of gravitational degrees of freedom is smaller than the number of polarizations, a result explained by the presence of a linear relationship between the scalar breathing and longitudinal modes \cite{Dyadina2022}. In addition, it has been established that the speed of gravitational waves in HMPG equals the speed of light, in full agreement with experimental data \cite{Dyadina2022}.

HMPG has also been investigated in the context of binary pulsars as a specific case of Horndeski theory. In \cite{Dyadina2018}, an expression for the energy loss due to gravitational radiation was derived, along with transition functions connecting the two theories. This study led to constraints on the background value and mass of the scalar field, based on observational data on orbital period decay from the systems PSR J0737-3039 and PSR J1738+0333. Recently, improved observational data from the PSR J0737-3039 system have been published \cite{Kramer_2021}. Therefore, one approach to improve the constraints on the parameters of HMPG is to apply the methods of \cite{Dyadina2018} using the updated data from \cite{Kramer_2021}, which is one of the objectives of this study.

Another goal of this work is to compute the gravitational waveforms emitted by inspiralling compact binary systems on quasicircular orbits in HMPG within the lowest post-Newtonian approximation. By applying the stationary phase approximation, we obtain their Fourier transforms and derive correction terms in the amplitude and phase of the gravitational waves relative to the corresponding results in GR. We also identify post-Einsteinian parameters in HMPG and find possible constraints on the background value of the scalar field, considering potential observations of gravitational waves by future ground-based detectors. This study represents an initial step in investigating gravitational waveforms within HMPG, aiming to refine constraints on this theory and deepen our fundamental understanding of gravity.

This article is organized into seven sections. The first and last are the introduction and conclusion, respectively. Section \ref{sec2} provides a description of HMPG and its scalar-tensor representation. In Section \ref{sec3}, we present the evolution of a binary system. Section \ref{sec4} outlines the calculation of gravitational waveforms in HMPG. Section \ref{spa} is devoted to the derivation of the Fourier transforms of the tensor amplitudes using the stationary phase approximation. In Section \ref{ppe}, we obtain post-Einsteinian parameters in HMPG and find possible constraints on the background value of the scalar field. Finally, Section \ref{concl} summarizes our findings.

Throughout this paper the Greek indices $(\mu, \nu,...)$ run over $0, 1, 2, 3$ and the signature is  $(-,+,+,+)$. We use natural units $c=\hbar=1$.

\section{Hybrid metric-Palatini gravity}\label{sec2}
	The action of HMPG combines the Einstein-Hilbert term with a Palatini contribution expressed as a general analytic function of the Palatini curvature $\Re$. It takes the following form \cite{Harko2012, Capozziello2015}: 
	\begin{equation}
		\label{action1}
		S=\frac{1}{2k^2}\int d^4x\sqrt{-g}[R+f(\Re)]+S_m,
	\end{equation}
where $k^2=8\pi G$, $G$ is the gravitational constant,  $g=det\{g_{\mu\nu}\}$ is the determinant of the metric, $R$ and $\Re$ are the metric and Palatini curvatures, respectively, and $S_m$~ is the matter action.  All deviations from GR are included in Palatini part. It is important to emphasize that in contrast to the metric approach, where the curvature depends only on the metric, in the Palatini approach, the affine connection is treated as an independent variable.

The simplest and most convenient way to study the HMPG theory is to present it in scalar-tensor form. After several transformations (for details, see \cite{Harko2012, Capozziello2015}) the action of HMPG can be represented as follows:
	\begin{equation}
		\label{action4}
		S=\frac{1}{2k^2}\int d^4x\sqrt{-g}\left[(1+\phi) R+\frac{3}{2\phi}\partial_\mu\phi\partial^\mu\phi-V(\phi)\right]+S_m,
	\end{equation}
where $\phi$ is a scalar field and $V(\phi)$ is a scalar potential. It should be noted that the action of the hybrid metric-Palatini gravity in the form of Eq.~\eqref{action4} can be transformed into a massive Brans--Dicke theory with an appropriate potential by means of a conformal redefinition of the scalar field as $\phi' = (1+\phi)/G$ \cite{Bronnikov_2019}. As a result, a Brans-Dicke type theory with a specific non-trivial coupling function is obtained \cite{Capozziello2015}:
\begin{equation}\label{wbd}
\omega_{BD} = -\frac{3}{2} \frac{\phi'}{1 - \phi'},
\end{equation}
which generalizes the well-known limits of Palatini $f(R)$ gravity ($\omega_{BD} = -3/2$) and metric $f(R)$ gravity ($\omega_{BD} = 0$). Despite the mathematical possibility of such conformal transformations, the question of the physical equivalence of the two approaches remains non-trivial and requires further investigation.

Further it is possible to obtain the field equations in the scalar-tensor form varying the action (\ref{action4}) with respect to the metric and scalar field. Thus the field equations are formulated as follows \cite{Harko2012, Capozziello2015}:
	\begin{eqnarray}
		\label{feq_g}
		&\frac{1}{1+\phi}\Bigg[k^2\left(T_{\mu\nu}-\frac{1}{2}g_{\mu\nu}T\right)+\frac{1}{2}g_{\mu\nu}\left(V+\nabla_\alpha\nabla^\alpha\phi\right)+\nabla_\mu\nabla_\nu\phi \nonumber\\
		&-\frac{3}{2\phi}\partial_\mu\phi\partial_\nu\phi\Bigg]=R_{\mu\nu},
\end{eqnarray}
\begin{equation}
		\label{feq_phi}
		-\nabla_\mu\nabla^\mu\phi+\frac{1}{2\phi}\partial_\mu\phi\partial^\mu\phi+\frac{\phi[2V-(1+\phi)V_\phi]}{3}=\frac{\phi k^2}{3}T,
	\end{equation}
 where $T_{\mu\nu}$ is the energy-momentum tensor, $T$ is its trace.
 
From this point onward,  the HMPG theory will be considered in the scalar-tensor form.

\section{Evolution of binary systems}\label{sec3}
The aim of this work is to calculate gravitational waveforms in HMPG up to the lowest post-Newtonian order. For this goal, we consider the dynamics of a binary system consisting of two compact objects.  Such systems lose energy through gravitational radiation. This process has previously been described in detail within the framework of Horndeski theory in \cite{Dyadina2018}, where HMPG was treated as a special case. In the present study, we make use of the results from \cite{Dyadina2018}, taking into account the corresponding transition functions between Horndeski gravity and HMPG (see Eq. (88) in \cite{Dyadina2018}).  It is also worth noting that in the previous work, all results for HMPG were obtained without considering the sensitivities. In this study, we address this shortcoming and include the sensitivities in our analysis.

The concept of sensitivities was first introduced by Eardley \cite{PhysRevD.8.3308}, who studied gravitational radiation from binary pulsars within the Brans-Dicke theory. Eardley proposed that the inertial mass is determined not only by the baryonic mass of the body but also by the energy of its gravitational binding, which, in turn, is influenced by an additional scalar field. As a result, the inertial mass should be treated as a function of the scalar field, affecting both the conservative and dissipative dynamics of a binary system.

The most general approach to calculating sensitivities in modified theories of gravity was developed in \cite{Barausse_2015}. The authors showed that sensitivities appear in the structure of all theories involving extra fields. Moreover, even in theories where such fields are not present explicitly (e.g., $f(R)$-gravity), sensitivities must still be considered, since these theories can be recast into the form of "GR plus an additional field" through mathematical transformations. HMPG is one such theory, as it can be represented in scalar-tensor form. Therefore, accounting for sensitivities is essential when studying strong-field gravitational phenomena.

To study the dynamics of a binary system, we assume that far from the source, the metric and the scalar field take the following forms:
\begin{equation}
g_{\mu\nu}=\eta_{\mu\nu}+h_{\mu\nu}, \ \ \ \ \phi=\phi_0+\varphi,
\end{equation}
where $\eta_{\mu\nu}$ is the Minkowski metric with a small perturbation $h_{\mu\nu}$, $\phi_0$ is the background value of the scalar field, $\varphi$ is its perturbation.

The matter action for a system of point-like masses can be written as
\begin{equation}\label{matter_action}
S_m=-\sum_a\int m_a(\phi)d\tau_a,
\end{equation}
where $m_a(\phi)$ are inertial masses of particles labeled by $a$ and $\tau_a$ is the proper time of the particle $a$ measured along its world-line $x^{\mu}_a$. From (\ref{matter_action}) it is clear that the mass $m_a(\phi)$  depends on position, since $\phi$ is a spacetime-dependent scalar field. The stress-energy tensor derived from the action (\ref{matter_action}) and its trace are given by
\begin{eqnarray}\label{stress-energy}
T^{\mu\nu}&=& \frac{1}{\sqrt{-g}}\sum_a m_a(\phi)\frac{u^{\mu}u^{\nu}}{u^0}\delta^3(\mathbf{r}-\mathbf{r}_a(t)), \nonumber\\
T&\equiv& g_{\mu\nu}T^{\mu\nu}=-\frac{1}{\sqrt{-g}}\sum_a \frac{m_a(\phi)}{u^0}\delta^3(\mathbf{r}-\mathbf{r}_a(t)),
\end{eqnarray}
where $u^{\mu}=d x^{\mu}_a/d \tau_a$ is four-velocity of the $a$-th particle,  $u_{\mu}u^{\mu}=-1$, and $\delta^3(\mathbf{r}-\mathbf{r}_a(t))$ is the three-dimensional Dirac delta function.

In matter action (\ref{matter_action}) the inertial mass is treated as an arbitrary function of the scalar field, which can be expanded in a Taylor series around the background value $\phi_0$:
\begin{equation}\label{inertial_mass} 
m_a(\phi)=m_a(\phi_0)\biggl[1+s_a\varphi+\frac{1}{2}\varphi^2(s_a^2+s'_a)+O(\varphi^3)\biggr].
\end{equation} 
Hereafter, we denote $m_a(\phi_0)$ as $m_a$, representing the inertial mass evaluated at the asymptotic scalar field value.  The quantities $s_a$ and $s_a'$ are the "first and second sensitivities":
\begin{equation}\label{sensitivities} 
s_a\equiv \frac{\partial(\ln m_a)}{\partial( \phi)}\bigg|_{\phi_0},\ \ \ s_a'\equiv \frac{\partial^2(\ln m_a)}{\partial( \phi)^2}\bigg|_{\phi_0}.
\end{equation} 

It is important to note that we define the sensitivities somewhat differently from the work \cite{PhysRevD.8.3308}. Specifically, we take the derivative with respect to the scalar field itself rather than its logarithm, following the approach of the work \cite{will1993theory}. The reason for this choice is that, in the form given by \eqref{sensitivities}, the sensitivities in HMPG are proportional to \(\Phi_s\), where \(\Phi_s\) is the surface gravitational potential of the body. In contrast, if the derivative were taken with respect to the logarithmic field, the sensitivities would be proportional to the product of the surface potential and the scalar field. In the case of Brans-Dicke theory, this distinction is insignificant since \(\phi \sim 1\). However, in HMPG, where \(\phi \to 0\), the sensitivities would significantly deviate from conventional values. Moreover, in HMPG, the scalar field \(\phi\) is dimensionless, eliminating the need to take the logarithm of the denominator, as is customary in Brans-Dicke theory.

Further, it is convenient to introduce the quantities $\theta_{\mu\nu}$ and $\theta$, which are defined as follows:
\begin{gather}
\begin{split}
\theta_{\mu\nu}&=h_{\mu\nu}-\frac{1}{2}\eta_{\mu\nu}h-\eta_{\mu\nu}\frac{\varphi}{\phi_0+1},\\
\theta&=-h-4\frac{\varphi}{\phi_0+1}.
\end{split}
\end{gather}

Employing the transverse gauge, $\partial_\mu\theta^{\mu\nu}=0$ simplifies the field Eqs.~\eqref{feq_g} and \eqref{feq_phi} in this way: 
\begin{equation}\label{eq}
\Box\theta_{\mu\nu}=-\frac{2k^2}{\phi_0+1}T_{\mu\nu},
\end{equation}
\begin{equation}\label{eqs}
\Box \varphi-m_\phi^2\varphi=-\frac{k^2\phi}{3}[T-2T_{,\phi}(1+\phi)],
\end{equation}
where $m_\phi$ is the scalar field mass, defined as $m_\phi^2=[2V_0-V'-(1+\phi_0)\phi_0V'']/3$. Here, the prime denotes differentiation with respect to the scalar field. The stress-energy tensor and its trace can be expressed as
\begin{eqnarray}\label{stress-energy_1}
	&T^{\mu\nu}=\sum_a m_au^{\mu}u^{\nu}\biggl(1-\frac{h^k_k}{2}-\frac{v_a^2}{2}+s_a\varphi\biggr)\delta^3(\mathbf{r}-\mathbf{r}_a(t)), \nonumber\\
	&T=-\sum_a m_a\biggl(1-\frac{h^k_k}{2}-\frac{v_a^2}{2}+s_a\varphi\biggr)\delta^3(\mathbf{r}-\mathbf{r}_a(t)),\nonumber\\
		&T_{,\phi}=-\sum_a m_a\biggl[s_a\biggl(1-\frac{h^k_k}{2}-\frac{v_a^2}{2}\biggr)+(s_a^2+s'_a)\varphi\biggr]\delta^3(\mathbf{r}-\mathbf{r}_a(t)),\nonumber\\
		&
	\end{eqnarray}
	where  $v_a$ and $m_a$ are the velocity and mass of the $a$-th particle, respectively.
	
	The leading-order static solution for the scalar field can be found from Eq.~\eqref{eqs} \cite{Dyadina2018}:
	\begin{equation}\label{phi} 
\varphi=-\frac{2\phi_0}{3}\sum_a\frac{m_a}{r_a}\biggl(1-2s_a(\phi_0+1)\biggr)e^{-m_\varphi r_a},
\end{equation} 
where $r_a=|\mathbf{r}-\mathbf{r}_a(t)|$.

Further, the solution to the tensor field equations~(\ref{eq}) within the 1PN approximation in the near zone is
\begin{eqnarray}\label{h} 
\theta_{00}&=&\frac{4}{\phi_0+1}\sum_a\frac{m_a}{r_a}, \nonumber\\
\theta_{ij}&=&\frac{4v_iv_j}{\phi_0+1}\sum_a\frac{m_a}{r_a}, \nonumber\\
\theta&=&-\frac{4}{\phi_0+1}\sum_a\frac{m_a}{r_a}.
\end{eqnarray} 
Thus, the leading order of metric perturbation is defined as \citep{Dyadina2018}:
\begin{eqnarray}\label{h} 
h_{00}&=&\frac{2}{\phi_0+1}\biggl[\sum_a\frac{m_a}{r_a}-\frac{\phi_0}{3} \sum_a\frac{m_a}{r_a}e^{-m_\varphi r_a}\biggl(1-2s_a(\phi_0+1)\biggr)\biggr],\nonumber\\
h_{ij}&=&\frac{2\delta_{ij}}{\phi_0+1}\biggl[\sum_a\frac{m_a}{r_a}+\frac{\phi_0}{3} \sum_a\frac{m_a}{r_a}e^{-m_\varphi r_a}\biggl(1-2s_a(\phi_0+1)\biggr)\biggr], \nonumber\\
h&=&\frac{4}{\phi_0+1}\biggl[\sum_a\frac{m_a}{r_a}+\frac{2\phi_0}{3}\sum_a\frac{m_a}{r_a}e^{-m_\varphi r_a} \biggl(1-2s_a(\phi_0+1)\biggr)\biggr],\nonumber\\
&
\end{eqnarray} 
with $h_{oi}=0$ in the leading order. Here $\delta_{ij}$ is the Kronecker delta.
	
	In a binary system, the compact objects can be treated as point-like bodies with masses $m_1$ and $m_2$ located at positions $\mathbf{r}_1$ and $\mathbf{r}_2$, respectively. However, it is convenient to reduce the analysis of such a system to an equivalent one-body system with  
	\begin{eqnarray}\label{onebody}
	\mu = \frac{m_1m_2}{m_1 + m_2}, \ \ \ \ m=m_1+m_2, \nonumber\\
	\mathbf{r}=\mathbf{r}_2-\mathbf{r}_1, \ \ \ \ r=|\mathbf{r}|,
\end{eqnarray} 
where $\mu$ is the reduced mass, $m$ is the total mass of the system, $\mathbf{r}$ is the relative coordinate.
	The corresponding equations of motion up to Newtonian order are given by:
\begin{eqnarray}\label{aa} 
\frac{d^2\mathbf{r}}{dt^2}=-\frac{\mathcal{G}m}{r^3}\mathbf{r}
\end{eqnarray} 
with the effective gravitational constant
\begin{eqnarray}\label{geff} 
&\mathcal{G}=\frac{k^2}{8\pi(1+\phi_0)}\biggl[1-\frac{\phi_0}{3}(1+m_\phi r)e^{-m_\phi r}\biggl(1-2s_a(\phi_0+1)\biggr)\nonumber\\
&\times\biggl(1-2s_b(\phi_0+1)\biggr)\biggr].
\end{eqnarray}

We now analyze the orbital dynamics of a quasi-circular binary system containing compact objects. Kepler’s third law takes the following form \cite{Dyadina2018}:
\begin{equation}\label{omega} 
\omega=\biggl(\frac{\mathcal{G}m}{r^3}\biggr)^{1/2},
\end{equation} 
where $\omega$ is the orbital frequency, related to the orbital period $P_b$ by $P_b=2\pi/\omega$. The orbital binding energy of such a system is given by:
\begin{equation}\label{energy} 
E=-\frac{\mathcal{G}m\mu}{2r}.
\end{equation}

The most significant dissipative effect is the  orbital period change due to the emission of GWs. The expression for loss of energy can be obtained from the first derivative of the orbital period, using Eqs.~(\ref{omega}) and (\ref{energy}):
\begin{equation}\label{energy_losses} 
\frac{\dot E}{E}=-\frac{2}{3}\frac{\dot P_b}{P_b}=\frac{2}{3}\frac{\dot \omega}{\omega}.
\end{equation} 

Using the results for energy loss from quasi-circular binary systems obtained within the framework of the Horndeski theory \cite{Dyadina2018}, we can find the expression for an average energy flux in HMPG. This result differs from that presented in \cite{Dyadina2018}, as we now explicitly take into account the presence of sensitivities in the theory.

It is important to note that energy loss can be divided into two parts: tensor and scalar. The average energy flux radiated in GWs from the tensor sector has the following form:
\begin{equation}\label{gravitational flux3} 
\left<\dot E_g\right> =-\frac{4k^2\mu^2(\mathcal{G}m)^3}{5\pi (\phi_0+1)r^5}.
\end{equation}

The total power of the gravitational radiation is \cite{Dyadina2018}
\begin{eqnarray}\label{total} 
\left<\dot E\right>=&-&\frac{4k^2\mu^2(\mathcal{G}m)^3}{5\pi (\phi_0+1
) r^5}\biggl[1- \frac{5\phi_0 r}{288\mathcal{G}m}A_d^2(v_{\varphi}(\omega))^3\Theta(\omega-m_\varphi)\nonumber\\
&-&\frac{5\phi_0\mu}{144\mathcal{G}m}A_d\bar{A}_{d}(v_{\varphi}(\omega))^3\Theta(\omega-m_\varphi)\nonumber\\
&-&\frac{\phi_0}{18}A_q^2 (v_{\varphi}(2\omega))^5\Theta(2\omega-m_\varphi)\nonumber\\
&+&\frac{\phi_0}{576}A_dA_o(v_{\varphi}(\omega))^5\Theta(\omega-m_\varphi)\biggr].
\end{eqnarray} 
where $v_{\varphi}(2\omega)=\sqrt{1-m_\phi^2 /4\omega^2}$ is the propagation speed of the scalar gravitational radiation, $\Theta(2\omega-m_\phi)$ is the Heaviside function.
We use the following definitions:
\begin{eqnarray}\label{Ad}
A_d=&&2(\phi_0+1)(s_2-s_1),\nonumber\\
A_q=&&1-2(\phi_0+1)\frac{s_2m_1+s_1m_2}{m},\nonumber\\ 
A_o=&&\frac{m_1-m_2}{m}-2(\phi_0+1)\frac{s_2m_1^2-s_1m_2^2}{m^2}
\end{eqnarray} 
\begin{eqnarray}\label{adbar}
\bar{A}_{d}=&-&\frac{7}{2(\phi_0+1)}\biggl(\frac{m_2}{m_1}-\frac{m_1}{m_2}\biggr)+7\biggl(\frac{m_2s_1}{m_1}-\frac{m_1s_2}{m_2}\biggr)\nonumber\\
&+&6(s_1-s_2)-\frac{2\phi_0(s_1-s_2)}{3}-\frac{23}{24}\frac{\phi_0}{(\phi_0+1)}\biggl(\frac{m_2}{m_1}-\frac{m_1}{m_2}\biggr)\nonumber\\
&-&\frac{5\phi_0}{4}\biggl(\frac{m_1s_2}{m_2}-\frac{m_2s_1}{m_1}\biggr)-2\phi_0\biggl(\frac{m_1s_1}{m_2}-\frac{m_2s_2}{m_1}\biggr)\nonumber\\
&-&\frac{7(\phi_0+1)s_1s_2\phi_0}{3}\biggl(\frac{m_2}{m_1}-\frac{m_1}{m_2}\biggr)\nonumber\\
&-&\phi_0\frac{(s_1+s_2)}{12}\biggl(\frac{m_2}{m_1}-\frac{m_1}{m_2}\biggr)+\frac{4(\phi_0+1)\phi_0}{3}(s_1'-s_2')\nonumber\\
&-&\frac{3(\phi_0+1)\phi_0}{2}\biggl(\frac{s^2_2m_1}{m_2}-\frac{s^2_1m_2}{m_1}\biggr)\nonumber\\
&-&\frac{8(\phi_0+1)^2\phi_0}{3}(s_1s_2'-s_2s_1')-\frac{4(\phi_0+1)\phi_0}{3}(s_2^2-s_1^2)\nonumber\\
&-&3(\phi_0+1)^2\phi_0\biggl(\frac{s^2_1s_2m_2}{m_1}-\frac{s^2_2s_1m_1}{m_2}\biggr)\nonumber\\
&-&\frac{8(\phi_0+1)^2\phi_0}{3}(s^2_1s_2-s^2_2s_1)\nonumber\\
&+&\frac{4(\phi_0+1)\phi_0}{3}\biggl(\frac{m_2s_1'}{m_1}-\frac{m_1s_2'}{m_2}\biggr)\nonumber\\
&-&\frac{8(\phi_0+1)^2\phi_0}{3}\biggl(\frac{s_1s_2'm_1}{m_2}-\frac{s_2s_1'm_2}{m_1}\biggr).
\end{eqnarray}

Due to the loss of energy to gravitational radiation, the orbital frequency $\omega$ increases. Using Eqs.~\eqref{omega} and \eqref{total}, we can calculate the leading order time derivative of the orbital frequency:
\begin{eqnarray}\label{omegadt}
\dot \omega=&&\frac{12k^2\mu(\mathcal{G}m)^{2/3}\omega^{11/3}}{5\pi (\phi_0+1
) }\nonumber\\
&\times&\biggl[1- \frac{5\phi_0 }{288(\mathcal{G}m\omega)^{2/3}}A_d^2(v_{\varphi}(\omega))^3\Theta(\omega-m_\varphi)\nonumber\\
&-&\frac{5\phi_0\mu}{144\mathcal{G}m}A_d\bar{A}_{d}(v_{\varphi}(\omega))^3\Theta(\omega-m_\varphi)\nonumber\\
&-&\frac{\phi_0}{18}A_q^2 (v_{\varphi}(2\omega))^5\Theta(2\omega-m_\varphi)\nonumber\\
&+&\frac{\phi_0}{576}A_dA_o(v_{\varphi}(\omega))^5\Theta(\omega-m_\varphi)\biggr].
\end{eqnarray} 
This result is essential for deriving the frequency-domain gravitational waveforms.

Previously, in work \cite{Dyadina2018},  the orbital period change was used to impose restrictions on the background value of the scalar field in HMPG. The parameters of the binary system PSR J0737-3039 were taken as observational data. Since then, the observational accuracy of the parameters for this binary system has significantly improved \cite{Kramer_2021}. Therefore, building upon the methods and results of  \cite{Dyadina2018}, we derive updated constraints on the parameters of the theory.

The system PSR J0737-3039  consists of two neutron stars, both observable as pulsars. The extraordinary closeness of the system components, small orbital period and the fact that we see almost edge-on system allow to investigate the manifestation of relativistic effects with the highest available precision. The updated observational data for this system is listed in the Table \ref{tab:J0737-3039}.

\begin{table}
		
			\caption{Parameters PSR J0737-3039 \citep{Kramer_2021}}\label{tab:J0737-3039}
			\begin{tabular}{l l l} 
				\hline
				
				Parameter & Physical meaning & Experimental value\\
				\hline
				$P_b $ & orbital period & $ 0.1022515592973(10)$ day\\

				$e$ & eccentricity & $ 0.087777023(61) $\\

				$\dot P_b^{obs}$ & secular change of & $ -1.247920(78) \times 10^{-12}$\\
				& the orbital period &\\
				
				$\dot P_b^{obs}/\dot P_b^{GR}$ & relation between $\dot P_b^{obs}$& $ 0.999963(63) $\\
				& and $\dot P_b^{GR}$ &\\
				
				$m_1 $ & mass of the first pulsar & $1.338185(+12/-14)\  m_{\bigodot}$\\
				
				$m_2 $ & mass of the second pulsar & $1.248868(+13/-11)\  m_{\bigodot}$\\
				
				$m $ & total system mass & $2.587052(+9/-7)\  m_{\bigodot}
				$\\
				\hline
				
			\end{tabular}
	
	\end{table}
	
The calculation method is to compare the predicted ratio \( \dot{P}_b^{\mathrm{th}} / \dot{P}_b^{\mathrm{GR}} \) with the observed ratio \( \dot{P}_b^{\mathrm{obs}} / \dot{P}_b^{\mathrm{GR}} \) at the 95\% confidence level:
\begin{equation}\label{deviation} 
\left| \frac{\dot{P}_b^{\mathrm{th}}}{\dot{P}_b^{\mathrm{GR}}} - \frac{\dot{P}_b^{\mathrm{obs}}}{\dot{P}_b^{\mathrm{GR}}} \right| \leq 2\sigma,
\end{equation}
where \( \sigma \) is the observational uncertainty. The expression for $\dot P_b^{th}/\dot P_b^{GR}$ in HMPG can be found using Eq.~\eqref{omegadt} and relation $P_b=2\pi/\omega$:
\begin{eqnarray}\label{first} 
\frac{\dot P_b^{th}}{\dot P_b^{GR}}=&&\frac{\mathcal{G}^{2/3}}{G^{2/3}(\phi_0+1)}\nonumber\\
&\times&\biggl[1- \frac{5\phi_0 }{288(\mathcal{G}m\omega)^{2/3}}A_d^2(v_{\varphi}(\omega))^3\Theta(\omega-m_\varphi)\nonumber\\
&-&\frac{5\phi_0\mu}{144\mathcal{G}m}A_d\bar{A}_{d}(v_{\varphi}(\omega))^3\Theta(\omega-m_\varphi)\nonumber\\
&-&\frac{\phi_0}{18}A_q^2 (v_{\varphi}(2\omega))^5\Theta(2\omega-m_\varphi)\nonumber\\
&+&\frac{\phi_0}{576}A_dA_o(v_{\varphi}(\omega))^5\Theta(\omega-m_\varphi)\biggr],
\end{eqnarray} 
where $\dot P_b^{GR}$ is the value of orbital decay predicted by GR:
\begin{equation}\label{GR} 
\dot P_b^{GR}=-\frac{192\pi\mu}{5m}\biggl(\frac{2\pi Gm}{P_b}\biggr)^{\frac{5}{3}}.
\end{equation} 

To correctly evaluate the Eq.~\eqref{first}, we must consider two limiting cases: $m_\varphi\ll\omega$ and $m_\varphi\gg\omega$. Let us begin with the first case.

When $m_\varphi\ll\omega$, the scalar field is very light. The Heaviside function takes the value of one, and all terms in the Eq.~\eqref{first} non-vanishing. Additionally, in the case of a light scalar field, we take $v(\omega)=v(2\omega)=1$. The expression for the effective gravitational constant contains both the mass of the scalar field and the distance scale specific to the system under consideration. It was previously shown that the distance at which the effects of the scalar field mass become significant exceeds the size of local astrophysical systems \cite{Leanizbarrutia2017}. Therefore, we can assume $m_\phi r\ll1$, and after substituting the Eq.~\eqref{geff} into Eq.~\eqref{first}, we obtain:
	\begin{eqnarray}\label{ns1}
	&\biggl|0.999963-\cfrac{1}{(\phi_0+1)^{5/3}}\biggl(1-\cfrac{\phi_0(1-2s_{NS}(1+\phi_0))^2}{3}\biggr)^{2/3}\nonumber\\
	&\times\biggl[1-\cfrac{\phi_0(1-2s_{NS}(1+\phi_0))^2}{18}\biggr]\biggr|\leq 0.000126.
	\end{eqnarray} 
Here we can estimate the sensitivities as $s_1\approx s_2\approx 0.2$ since both objects are neutron stars. In this case, dipole parameter vanishes, $A_d=0$, because $s_1=s_2$. Thus, only the quadrupole contribution remains in Eq.~\eqref{ns1} for the scalar sector.  Consequently, we find the following constraints on the background value of the scalar field:
\begin{equation}\label{phins} 
	-5\times10^{-5}<\phi_0<9.2\times10^{-5}.
	\end{equation}

We conclude that the constraints derived from the double pulsar are comparable in magnitude to those obtained from the "Cassini" experiment \cite{Cassini2003}. While these constraints are slightly less stringent than those obtained from the Solar System ($|\phi_0|<3.4\times10^{-5}$) \cite{Leanizbarrutia2017, Dyadina2019}. Continued observations of the system PSR J0737-3039 will further refine the limits on $\phi_0$. Additionally, performing a full post-Keplerian test may further improve existing constraints.

Next we consider the case $m_\varphi\gg\omega$. In this regime, all terms in Eq.~\eqref{first} containing the Heaviside function vanish. The additional term in the square brackets of the expression for the effective gravitational constant also disappears due to $e^{m_\varphi r}\to0$. Substituting the Eq.~\eqref{geff} into Eq.~\eqref{first} and applying the method \eqref{deviation}, we obtain:
	\begin{eqnarray}\label{ns}
	&&\biggl|0.999963-\cfrac{1}{(\phi_0+1)^{5/3}}\biggr|\leq 0.000126.
	\end{eqnarray} 
 Finally, we find the following constraints on the background value of the scalar field:
	\begin{equation}\label{phins} 
	-5.3\times10^{-5}<\phi_0<9.8\times10^{-5}.
	\end{equation}
	
	Let us emphasize that this restriction is valid only in the limit $m_\varphi\gg\omega$. This limit was previously studied in the work \cite{Leanizbarrutia2017}. In this paper, the authors were able to establish constraints on the theory using experimental data for the gravitational constant in the case where the mass of the scalar field is much larger than the inverse dimensions of the local system. They obtained that $|\phi_0|<5\times10^{-4}$. Thus, the constraints derived in this work, as given in Eq.~\eqref{phins}, represent an improvement over those obtained in Ref.\cite{Leanizbarrutia2017}.

\section{Gravitational waveforms}\label{sec4}

To obtain gravitational waveforms in HMPG, it is necessary to solve the Eqs.~\eqref{eq} and \eqref{eqs} in far zone. The solution consists of tensor and scalar parts. In \cite{Dyadina2018}, the linearized field equations were solved in the far zone using the Green's function method. The metric perturbation is formulated in terms of mass multipole moments, while the scalar field is described using scalar multipole moments. Analogous to GR, we consider the metric perturbation only up to the quadrupole order. The  quantity $\theta_{ij}$ that describes the tensor radiation was obtained in the work \cite{Dyadina2018}:
\begin{equation}\label{spatial solution2} 
\theta_{ij} =\frac{k^2}{4\pi R(\phi_0+1)}\frac{\partial^2}{\partial t^2}\sum_a m_ar^{i}_ar^{j}_a,
\end{equation} 
where $R$ represents the coordinate distance from the compact binary to the observer.

After differentiation and using Eq.~\eqref{aa}, expression \eqref{spatial solution2} is reduced to the form:
\begin{equation}\label{spatial solution3} 
\theta_{ij} =\frac{k^2\mu}{2\pi R(\phi_0+1)}\biggl(v^{i}v^{j}-\frac{\mathcal{G}mr^ir^j}{r^3}\biggr),
\end{equation} 
where $v^i=\dot r^i=\dot r_2^i-\dot r_1^i$.

For the circular orbit at leading order $v^2=\mathcal{G}m/r$. Using this relation we can reduce the Eq.~\eqref{spatial solution3} to the following form:
\begin{equation}\label{spatial solution4} 
\theta_{ij} =\frac{k^2\mathcal{G}m\mu}{2\pi Rr(\phi_0+1)}(\hat v^{i}\hat v^{j}-\hat r^i \hat r^j),
\end{equation} 
where $\hat v^{i}=v^i/v$, $ \hat r^i= r^i/r$.

Now let us consider scalar part. The scalar field Eq.~\eqref{eqs} up to  second order in perturbations can be expressed as
\begin{equation}\label{sourcce}
(\Box_\eta-m_\phi^2) \varphi=-k^2S
\end{equation}
with scalar source term
\begin{equation}\label{sour}
S=\frac{\phi_0}{3}\biggl(1-\frac{1}{2}\theta-\frac{\varphi}{\phi_0+1}+\frac{\varphi}{\phi_0}\biggr)(T-2T_{,\phi}(1+\phi_0+\varphi)).
\end{equation}
Here $\Box_\eta$ denotes the d'Alembertian in Minkowski spacetime.

The solution of Eq.~\eqref{sourcce} measured by an observer in the far zone can be separated into  a ``massless'' part, $\phi_B$ and ``massive''  part, $\phi_m$: \cite{Alsing_2012, Liu_2020, Higashino_2023}:
\begin{equation}\label{fi} 
\phi_B=\frac{k^2}{4\pi }\int\int\frac{S(t',\mathbf{r'})\delta(t-t'-|\mathbf{R}-\mathbf{r'}|)}{|\mathbf{R}-\mathbf{r'}|}dt'd^3\mathbf{r'},
\end{equation} 
\begin{equation}\label{fi1} 
\phi_m=-\frac{k^2}{4\pi }\int_Nd^3\mathbf{r}'\int_0^\infty \frac{J_1(z)S\bigl(t-\sqrt{|\mathbf{R}-\mathbf{r'}|^2+(\frac{z}{m_\phi})^2},\mathbf{r}'\bigr)}{\sqrt{|\mathbf{R}-\mathbf{r'}|^2+(\frac{z}{m_\phi})^2}}dz,
\end{equation} 
where $J_1$ is the Bessel function of the first kind, $S(t, r)$ is the source function from Eq.~\eqref{sour}, and $z=m_\phi\sqrt{(t-t')^2-|\mathbf{R}-\mathbf{r'}|^2}$.
In the far zone $(R\gg |r'|)$, it is possible to use  the approximation $|\mathbf{R}-\mathbf{r'}|=R-\mathbf{r'}\cdot \mathbf{n}$ with $\mathbf{n}=\mathbf{R}/R$. We also replace $t'$ with $t' =t-R+\mathbf{r'}\cdot \mathbf{n}$. Then, we perform multipole expansion of the time-dependent part of the scalar source term $S$. As a result, we obtain:
\begin{equation}\label{fi2} 
\phi_B=\frac{k^2}{4\pi R}\sum_{l=0}^\infty\frac{1}{l!}\frac{\partial^l}{\partial t^l}\int S(t-R,\mathbf{r'})(\mathbf{n} \cdot\mathbf{r}')^ld^3\mathbf{r'},
\end{equation} 
\begin{equation}\label{fi3} 
\phi_m=-\frac{k^2}{4\pi R}\sum_{l=0}^\infty\frac{1}{l!}\frac{\partial^l}{\partial t^l}\int (\mathbf{n} \cdot\mathbf{r}')^ld^3\mathbf{r'}\int_0^\infty \frac{J_1(z)S(t-Ru,\mathbf{r'})}{u^{l+1}}dz,
\end{equation} 
where
\begin{equation}\label{uz}
u=\sqrt{1+\frac{z^2}{m_\phi R^2}}.
\end{equation}
Next, we substitute the post-Newtonian expression for the source $S$ from Eq.~\eqref{sour}  into Eqs.~\eqref{fi2} and \eqref{fi3}. In this way, we obtain an expression for the gravitational waveform $\phi(t, R)$ in the far zone. Here, we use the  one-body approximation from Eq.~\eqref{onebody}. After performing the integration we save only the terms up to order $O(\frac{mv^2}{R})$ and $O(\frac{m^2}{Rr'})$ in the monopole (l=0) and quadrupole (l=2) contributions. As a result, we obtain
\begin{eqnarray}\label{phib}
\phi_B=&-&\frac{k^2\phi_0(\phi_0+1)\mu}{6\pi R}\biggl[\frac{k^2}{8\pi(\phi_0+1)}\frac{m}{r}\biggl((s_a+s_b)-\frac{1}{\phi_0+1}\biggr)\nonumber\\
&-&\frac{1}{4} \Gamma v^2+\frac{k^2\phi_0(\phi_0+1)}{12\pi}\frac{me^{-m_\phi r}}{r}F_s-(s_a-s_b)\mathbf{v}\cdot\mathbf{n}\nonumber\\
&+&\frac{1}{2}\Gamma(\mathbf{v}\cdot\mathbf{n})^2-\frac{1}{2}\frac{\mathcal{G}m}{r^3}\Gamma(\mathbf{r}\cdot\mathbf{n})^2\biggr]\biggl|_{t-R},
\end{eqnarray}
where
\begin{equation}
\Gamma=\frac{1}{(\phi_0+1)}-\frac{2(m_bs_a+m_as_b)}{m},
\end{equation}
\begin{eqnarray}
F_s=&&\frac{1-2s_b(\phi_0+1)}{2(\phi_0+1)}\biggl[4(s_a^2+s_a')+\frac{8s_a}{(\phi_0+1)}+\frac{4s_a}{\phi_0(\phi_0+1)}\nonumber\\
&-&\frac{1}{(\phi_0+1)^2}-\frac{2}{\phi_0(\phi_0+1)}\biggr]-\frac{1-2s_a(\phi_0+1)}{2(\phi_0+1)}\biggl[4(s_b^2+s_b')\nonumber\\
&+&\frac{8s_b}{(\phi_0+1)}+\frac{4s_b}{\phi_0(\phi_0+1)}-\frac{1}{(\phi_0+1)^2}-\frac{2}{\phi_0(\phi_0+1)}\biggr].
\end{eqnarray}
Also we find:
\begin{eqnarray}\label{phim}
\phi_m=&-&\frac{k^2\phi_0(\phi_0+1)\mu}{6\pi R}\biggl[\frac{k^2}{8\pi(\phi_0+1)}\biggl((s_a+s_b)-\frac{1}{\phi_0+1}\biggr)I_1\bigl[\frac{m}{r}\bigr]\nonumber\\
&-&\frac{1}{4}\Gamma I_1[v^2]+\frac{k^2\phi_0(\phi_0+1)}{12\pi}F_sI_1\bigl[\frac{me^{-m_\phi r}}{r}\bigr]-(s_a-s_b)I_3[\mathbf{v}\cdot\mathbf{n}]\nonumber\\
&+&\frac{1}{2}\Gamma I_3[(\mathbf{v}\cdot\mathbf{n})^2] -\frac{1}{2}\Gamma I_3\bigl[\frac{\mathcal{G}m}{r^3}(\mathbf{r}\cdot\mathbf{n})^2\bigr]\biggr]\biggl|_{t-Ru},
\end{eqnarray}
where
\begin{equation}
I_n[f(t)]=\int_0^\infty\frac{f(t-Ru)J_1(z)}{u^n}dz.
\end{equation}

A gravitational-wave detector measures the distance $\xi^i$ between freely moving test particles. If the separation between them is smaller than the wavelength of GW, and the test masses move slowly, the geodesic deviation equation reduces to $d^2\xi^i/dt^2=-R_{0i0j}\xi^j$, where $R_{0i0j}$ is the electric part of Riemann tensor \cite{maggiore2007gravitational}. Then the gravitational-wave field $\mathbf{h}_{ij}$ is defined by:
\begin{equation}\label{riemann}
R_{0i0j}=-\frac{1}{2}(\partial_0^2h_{ij}+\partial_i\partial_jh_{00})=-\frac{1}{2}\frac{d^2}{dt^2}\mathbf{h}_{ij},
\end{equation}
where we take into accaunt only linear terms. Under the transverse-traceless gauge Eq.~\eqref{riemann} takes the following form:
\begin{equation}\label{hij}
\partial_0^2\mathbf{h}_{ij}=\partial_0^2\theta_{ij}^{TT}-\frac{\delta_{ij}}{\phi_0+1}\partial_0^2\varphi+\frac{\partial_i\partial_j\varphi}{\phi_0+1},
\end{equation}
where ``TT'' represents the transverse-traceless gauge, $\delta_{ij}$ is the Kronecker delta. Taking into accaunt Eqs.~\eqref{phib} and \eqref{phim}, the scalar field $\varphi$ can be expressed as
\begin{equation}\label{varphi}
\varphi=\phi_B(t-R,\mathbf{n})+\phi_m(t-Ru,\mathbf{n}), 
\end{equation}
with
\begin{eqnarray}\label{phib1}
\phi_B=&&\frac{k^2\phi_0(\phi_0+1)\mu}{6\pi R}\biggl[(s_a-s_b)\mathbf{v}\cdot\mathbf{n}+\frac{\Gamma}{2}(\mathbf{v}\cdot\mathbf{n})^2\nonumber\\
&-&\frac{\Gamma}{2}\frac{\mathcal{G}m}{r^3}(\mathbf{r}\cdot\mathbf{n})^2\biggr],
\end{eqnarray}
\begin{equation}\label{phib1}
\phi_m=-\frac{k^2\phi_0(\phi_0+1)\mu}{6\pi R}\int_0^\infty dzJ_1(z)\psi_m,
\end{equation}
where
\begin{eqnarray}\label{phim1}
\psi_m=\biggl[(s_a-s_b)\frac{\mathbf{v}\cdot\mathbf{n}}{u^2}+\frac{\Gamma}{2}\frac{(\mathbf{v}\cdot\mathbf{n})^2}{u^3} -\frac{\Gamma}{2}\frac{\mathcal{G}m}{r^3}\frac{(\mathbf{r}\cdot\mathbf{n})^2}{u^3}\biggr]\biggr|_{t-Ru}.
\end{eqnarray}

There are no monopole terms in Eq.~\eqref{varphi}, since they do not contribute to the wavelike behavior of the scalar field perturbations. Therefore, here and further we do not consider them.

Equation \eqref{hij} contains, in addition to the time derivatives, the spatial derivatives of the scalar field. To solve this equation we need to perform the following transformations, following the approach of \cite{Liu_2018}:
\begin{eqnarray}\label{phi0}
\partial_i\partial_j\phi_B(t-R,\mathbf{n})=n_in_j\partial_0^2\phi_B(t-R,\mathbf{n})+O\biggl(\frac{1}{R^2}\biggr),\nonumber\\
\partial_i\partial_j\psi_m(t-Ru,\mathbf{n})=\frac{n_in_j}{u^2}\partial_0^2\psi_m(t-Ru,\mathbf{n})+O\biggl(\frac{1}{R^2}\biggr),
\end{eqnarray}
where $n_i=r_i/R$. Then Eq.~\eqref{hij} reduces to
\begin{eqnarray}\label{hij1}
\partial_0^2\mathbf{h}_{ij}=\partial_0^2\biggl[&&\theta_{ij}^{TT}-\frac{(\delta_{ij}-n_in_j)}{\phi_0+1}\varphi\nonumber\\
&-&n_in_j\frac{k^2\phi_0\mu}{6\pi R}\int_0^\infty dzJ_1(z)\biggl(\frac{1}{u^2}-1\biggr)\psi_m\biggr].
\end{eqnarray}

Metric theories of gravity can predict up to six polarization states of GWs \cite{PhysRevD.8.3308, will1993theory}. GR contains only ``plus'' and ``cross'' modes. In addition, gravitational theory can include a scalar breathing mode, a scalar longitudual mode, and two vectorial modes. HMPG consists of tensor and scalar fields, thus gravitational-wave field takes the following form:
\begin{eqnarray}\label{hmatrix}
\mathbf{h}_{ij}=
\begin{pmatrix}
  h_++h_b& h_\times&0\\
  h_\times& -h_++h_b&0\\
  0&0&h_L
\end{pmatrix}.
\end{eqnarray}
We consider the GWs propagating along the z direction in a three dimensional Cartesian coordinate system $(x, y, z)$. In this case $n_x =n_y =0$ and $n_z =1$. The presence of a nonminimally coupled scalar field in HMPG gives rise to both scalar breathing and longitudual modes. The latter appears only if the scalar field is massive \cite{Maggiore_2000}. The correspondence between the number of degrees of freedom and polarization modes in HMPG was previously investigated in work \cite{Dyadina2022}. It was shown that there exists a linear relationship between the scalar breathing and scalar longitudinal modes. This result is consistent with the presence of three dynamical degrees of freedom in HMPG.

Now, using Eqs.~\eqref{hij} and \eqref{hmatrix}, we can obtain the four polarization modes of GWs in HMPG:
\begin{equation}\label{h++}
h_+=-\frac{k^2\mathcal{G}^{2/3}(M_c)^{5/3}\omega^{2/3}}{2\pi R(\phi_0+1)}\frac{1+\cos^2i}{2}\cos(2\Phi),
\end{equation}
\begin{equation}\label{h*}
h_\times=-\frac{k^2\mathcal{G}^{2/3}(M_c)^{5/3}\omega^{2/3}}{2\pi R(\phi_0+1)}\cos i\sin(2\Phi),
\end{equation}
\begin{eqnarray}\label{hb}
h_b=&-&\frac{k^2\phi_0\mu}{6\pi R}\biggl[(s_a-s_b)v \sin{i}\cos(\Phi)-\frac{1}{2}\Gamma v^2\sin^2i\cos(2\Phi)\nonumber\\
&-&\int_0^\infty dzJ_1(z)\biggl(\frac{1}{u^2}(s_a-s_b)v \sin{i}\cos(\Phi)\nonumber\\
&-&\Gamma\frac{v^2}{2u^3}\sin^2i\cos(2\Phi)\biggr)\biggr],
\end{eqnarray}
\begin{eqnarray}\label{hL}
h_L=&-&\frac{k^2\phi_0\mu}{6\pi R}\biggl[\int_0^\infty dzJ_1(z)\biggl(\frac{1}{u^2}-1\biggr)\\
&\times&\biggl(\frac{1}{u^2}(s_a-s_b)v \sin{i}\cos(\Phi)-\Gamma\frac{v^2}{2u^3}\sin^2i\cos(2\Phi)\biggr)\biggr],\nonumber
\end{eqnarray}
where $\Phi(t)= \int_{t_0}^t \omega(t')dt'$ is the orbital phase of the binary system, $i$ is the inclination angle of the binary orbital angular momentum along the line of sight, and $M_c=\mu^{3/5}m^{2/5}$ is the chirp mass.  In addition, we introduce the following unit vectors:
\begin{equation}
\mathbf{n}=(0,\sin i, \cos i), \  \mathbf{\hat r}=(\cos \Phi,\sin \Phi, 0),\ \mathbf{\hat v}=(-\sin \Phi,\cos \Phi, 0).
\end{equation}

Further we take the asymptotic behavior of the integrals in $h_b$ and $h_L$ in the limit  $R\to \infty$. The details of this calculation are shown in \cite{Liu_2018}. Here ,we demonstrate only brief mathemathical derivation and the final result. We use the following notations:
\begin{equation}
I_1=\int_0^\infty\frac{\omega(t-Ru)^{1/3}J_1(z)}{u^2}\cos(\Phi(t-Ru))dz,
\end{equation}
\begin{equation}
I_2=\int_0^\infty\frac{\omega(t-Ru)^{2/3}J_1(z)}{u^3}\cos(2\Phi(t-Ru))dz,
\end{equation}
\begin{equation}
I_3=\int_0^\infty\frac{\omega(t-Ru)^{1/3}J_1(z)}{u^2}\biggl(\frac{1}{u^2}-1\biggr)\cos(\Phi(t-Ru))dz,
\end{equation}
\begin{equation}
I_4=\int_0^\infty\frac{\omega(t-Ru)^{2/3}J_1(z)}{u^3}\biggl(\frac{1}{u^2}-1\biggr)\cos(2\Phi(t-Ru))dz,
\end{equation}
where $u=\sqrt{1+\bigl(\frac{z}{m_\phi R}\bigr)^2}$ and $\omega(t)=d\Phi(t)/dt$. In the real situation we have $\omega\gg m_\phi$. Follow to \cite{Liu_2018} at leading order, we have $I_1$
\begin{eqnarray}
&I_1\simeq\omega(t-R)^{1/3}\cos(\Phi(t-R))-\omega(t-Ru_1)^{-2/3}\\
&\times\sqrt{\omega(t-Ru_1)^{2}-m_\phi^2}\cos\biggl(\frac{m_\phi^2R}{\sqrt{\omega(t-Ru_1)^{2}-m_\phi^2}}+\Phi(t-Ru_1)\biggr)\nonumber
\end{eqnarray}
with 
\begin{equation}\label{u1}
u_n=\frac{n\omega(t-R)}{\sqrt{n^2\omega(t-R)^2-m_\phi^2}}.
\end{equation}
Similarly, we have the asymptotic expression of $I_2$
\begin{eqnarray}
&I_2\simeq\omega(t-R)^{2/3}\cos(2\Phi(t-R))-\omega(t-Ru_2)^{2/3}\\
&\biggl(1-\frac{m_\phi^2}{4\omega(t-Ru_2)^2}\biggr)\cos\biggl(\frac{m_\phi^2R}{\sqrt{4\omega(t-Ru_2)^{2}-m_\phi^2}}+2\Phi(t-Ru_2)\biggr),\nonumber
\end{eqnarray}
\begin{equation}
I_3\simeq\frac{m_\phi^2}{\omega^{8/3}}\sqrt{\omega^2-m_\phi^2}\cos\biggl(\frac{m_\phi^2R}{\sqrt{\omega^{2}-m_\phi^2}}+\Phi\biggr)\biggl|_{t-Ru_1},
\end{equation}
\begin{equation}
I_4\simeq\frac{m_\phi^2}{4\omega^{4/3}}\biggl(1-\frac{m_\phi^2}{4\omega^2}\biggr)\cos\biggl(\frac{m_\phi^2R}{\sqrt{4\omega^{2}-m_\phi^2}}+2\Phi\biggr)\biggl|_{t-Ru_2}.
\end{equation}

Now, having obtained estimates for the integrals  $I_1$ and $I_2$ which appear in Eqs.~\eqref{hb} and \eqref{hL}, we can write the final expressions for $h_b$ and $h_L$:
\begin{equation}
h_b=h_{b1}+h_{b2}
\end{equation}
\begin{eqnarray}\label{hb1}
h_{b1}=&-\frac{k^2\mu\phi_0(s_a-s_b)}{6\pi R}(\mathcal{G}m)^{1/3}\sin{i}\ \ \omega^{1/3}\sqrt{1-\frac{m_\phi^2}{\omega^2}}\nonumber\\
&\times\cos{\biggl(\frac{m_\phi^2R}{\sqrt{\omega^2-m_\phi^2}}+\Phi\biggr)}\Theta(\omega-m_\phi)\biggl|_{t-Ru_1},
\end{eqnarray}
\begin{eqnarray}\label{hb2}
h_{b2}=&\frac{k^2\phi_0M_c^{5/3}}{12\pi R}\Gamma\mathcal{G}^{2/3}\sin^2i\ \ \omega^{2/3}\biggl(1-\frac{m_\phi^2}{4\omega^2}\biggr)\nonumber\\
&\times\cos{\biggl(\frac{m_\phi^2R}{\sqrt{4\omega^2-m_\phi^2}}+2\Phi\biggr)}\Theta(2\omega-m_\phi)\biggl|_{t-Ru_2},
\end{eqnarray}
and
\begin{equation}
h_L=h_{L1}+h_{L2}
\end{equation}
\begin{eqnarray}\label{hL1}
h_{L1}=&-\frac{m_\phi^2}{\omega^2}\frac{k^2\mu\phi_0(s_a-s_b)}{6\pi R}(\mathcal{G}m)^{1/3}\sin{i}\ \ \omega^{1/3}\sqrt{1-\frac{m_\phi^2}{\omega^2}}\nonumber\\
&\cos{\biggl(\frac{m_\phi^2R}{\sqrt{\omega^2-m_\phi^2}}+\Phi\biggr)}\Theta(\omega-m_\phi)\biggl|_{t-Ru_1},
\end{eqnarray}
\begin{eqnarray}\label{hL2}
h_{L2}=&\frac{m_\phi^2}{4\omega^2}\frac{k^2\phi_0M_c^{5/3}}{12\pi R}\Gamma\mathcal{G}^{2/3}\sin^2i\ \ \omega^{2/3}\biggl(1-\frac{m_\phi^2}{4\omega^2}\biggr)\nonumber\\
&\cos{\biggl(\frac{m_\phi^2R}{\sqrt{4\omega^2-m_\phi^2}}+2\Phi\biggr)}\Theta(2\omega-m_\phi)\biggl|_{t-Ru_2},
\end{eqnarray}
We have used the relation $v=(\mathcal{G}m\omega)^{1/3}$ and discarded terms of order $O(\frac{e^{-R}}{R})$. Due to the presence of the Heaviside function $\Theta$, a binary system can radiate scalar waves only if the orbital frequency $\omega$ is high enough. It is worth noting that the amplitudes labeled with index ‘1’ correspond to scalar dipole radiation, while those with the index ‘2’ correspond to scalar quadrupole radiation.

It is evident that a straightforward linear relation exists between the breathing $h_b$ and the longitudinal states $h_L$:
\begin{equation}
h_{L1}=\frac{m_\phi^2}{\omega^2}h_{b1},\ \ \ \ \ h_{L2}=\frac{m_\phi^2}{4\omega^2}h_{b2}.
\end{equation}
Using this result, we can estimate the ratios between the amplitudes of different polarization states. Let us consider a black hole-neutron star binary system with component masses $m_{BH}=5 M_\odot$ and $m_{SN}=1.4 M_\odot$, and sensitivities $s_{BH}\sim0.5$ and $s_{NS}\sim0.2$, respectively. We also assume that the scalar field has a mass $m_\phi\sim10^{-18}$ eV \cite{Dyadina2018}, the parameter $\phi_0$ is of order $10^{-5}$ \cite{Leanizbarrutia2017, Dyadina2019}, and the frequency of the tensor wave is approximately 100 Hz. Under these conditions,  the amplitude ratio of the scalar breathing mode to the tensor ``plus'' mode are given by:
\begin{equation}
\frac{h_{b1}}{h_+}\sim10^{-5},\ \ \ \ \frac{h_{b2}}{h_+}\sim10^{-6}
\end{equation}
 and the ratio between the longitudinal and breathing scalar modes is approximately
\begin{equation}
\frac{h_L}{h_b}\sim10^{-12}.
\end{equation}

Thus, we conclude that tensor radiation dominates. The current constraints on the free parameters of HMPG and the capabilities of modern gravitational-wave detectors make detecting scalar radiation a challenging task. It is worth noting that the smaller the value of the background scalar field considered, the wider the gap between the amplitudes of tensor radiation and the scalar breathing mode becomes.  Additionally, at frequencies comparable to the scalar field mass $\omega\gtrsim m_\phi$, the longitudinal component of scalar radiation can become comparable in magnitude to the scalar breathing mode. 

Furthermore, it should be noted that the expressions obtained for the amplitudes of gravitational waves Eqs.~\eqref{h++}, \eqref{h*}, \eqref{hb1}, \eqref{hb2}, \eqref{hL1}, and~\eqref{hL2} formally coincide with the corresponding results of Ref. \cite{Liu_2020} derived from massive Brans-Dicke theory via the conformal transformation $\phi' = (1+\phi)/G$ and the coupling function $\omega_{BD}$ defined in Eq.~\eqref{wbd}. However, it is important to note that we define the sensitivities of compact objects differently from Ref.~\cite{Liu_2020}, following the approach of Ref.~\cite{will1993theory} as demonstrated in Eq.~\eqref{sensitivities}, by taking the derivative with respect to the scalar field itself rather than its logarithm. This choice is motivated by the fact that sensitivities are generally proportional to the surface gravitational potential $\Phi_s$, but using a logarithmic derivative in HMPG would result in unphysically large scalar contributions, particularly since $\phi \rightarrow 0$. Therefore, the question of physical equivalence between the two approaches remains non-trivial and justifies an independent study of HMPG in the context of gravitational wave generation and detection.

\section{Frequency-domain gravitational waveforms}\label{spa}

When analyzing gravitational waveforms in comparison with observations, it is standard to apply a Fourier transformation to the polarization modes $h_+, h_\times, h_b, h_L$ with respect to the frequency $f$. As shown in the previous section, the amplitudes of $h_+$ and $h_\times$ are significantly larger than those of $h_b$ and $h_L$. Therefore, we focus on estimating deviations from GR in the Fourier domain for the dominant tensor polarizations $h_+$ and $h_\times$. Further we can use the stationary phase approximation (SPA) to calculate the Fourier transform. Applying of SPA becomes possible because during the inspiral the change in the orbital frequency over a single period is negligible. We now proceed with the Fourier transformation:
\begin{equation}
\tilde{h}_\alpha(f)=\int dt \ {h}_\alpha(t)e^{2i\pi ft},
\end{equation}
where $\alpha=+,\times$. For the $h_+$ mode, using Eq.~\eqref{h++}, we obtain
\begin{eqnarray}\label{h+++}
\tilde h_+(f)=&-\frac{k^2\mathcal{G}^{2/3}M_c^{5/3}(1+\cos^2i)}{8\pi R(\phi_0+1)}e^{2i\pi fR}\int dt \ \omega(t)^{2/3}\nonumber\\
&\times\bigl[e^{i(2\Phi(t)+2\pi ft)}+e^{i(-2\Phi(t)+2\pi ft)}\bigr].
\end{eqnarray}
The first term inside the square brackets does not possess a stationary point, that is, there is no value of $t$ that satisfies the condition $d[2\Phi(t)+2\pi ft]/dt=0$. Consequently, the first term in Eq.~\eqref{h+++} is always oscillating rapidly, and its contribution can typically be disregarded during integration. The stationary phase point $t_*$ of the second term is determined by:
\begin{equation}
\frac{d}{dt}[-2\Phi(t)+2\pi ft]\biggl|_{t=t_*}=0 \ \ \ \ \to \ \ \ \ \omega(t_*)=\pi f,
\end{equation}
it is taken into account that $\Phi(t)=\int_{t_0}^t\omega(t')dt'+\Phi_0$, where $\Phi_0$ is the initial phase at $t = t_0$.  We can expand $\Phi(t)$ around $t = t_*$ as $\Phi(t) = \Phi(t_*) + \pi f(t-t_*) +\dot\omega(t_*)(t-t_*)^2/2 + O((t-t_*)^3)$. Then we obtain
\begin{eqnarray}\label{h1}
\tilde h_+(f)=&-\frac{k^2\mathcal{G}^{2/3}M_c^{5/3}(1+\cos^2i)}{8\pi R(\phi_0+1)}e^{i(2\pi fR-2\Phi(t_*)+2\pi ft_*)}\nonumber\\
&\int dt \ \omega(t)^{2/3}e^{-i\dot\omega(t_*)(t-t_*)^2}.
\end{eqnarray}
Since $\int dt \ \omega(t)^{2/3}e^{-i\dot\omega(t_*)(t-t_*)^2}\simeq \omega(t_*)^{2/3}\frac{\sqrt{\pi}}{\sqrt{\dot\omega(t_*)}}e^{-i\pi/4}$, we find
\begin{equation}\label{h2}
\tilde h_+(f)=-\frac{k^2\mathcal{G}^{2/3}M_c^{5/3}(1+\cos^2i)}{8\pi R(\phi_0+1)}\omega(t_*)^{2/3}\frac{\sqrt{\pi}}{\sqrt{\dot\omega(t_*)}}e^{i\Psi_+},
\end{equation}
where
\begin{equation}
\Psi_+=2\pi ft_*-2\Phi(t_*)+2\pi fR-\pi/4.
\end{equation}
Similarly, the Fourier-transformed mode of $h_\times(f)$ is given by
\begin{equation}\label{h3}
\tilde h_\times(f)=-\frac{k^2\mathcal{G}^{2/3}M_c^{5/3}(\cos i)}{4\pi R (\phi_0+1)}\omega(t_*)^{2/3}\frac{\sqrt{\pi}}{\sqrt{\dot\omega(t_*)}}e^{i\Psi_\times},
\end{equation}
where
\begin{equation}
\Psi_\times=\Psi_++\pi/2.
\end{equation}

As stated in Eq.~\eqref{omegadt}, the orbital frequency $\omega$ increases over time. At a critical moment $t_c$, $\omega$ becomes sufficiently large, eventually reaching an infinite value, i.e., $\omega(t_c)\to\infty$. Under these circumstances, the time $t_*$ can be defined as:
\begin{equation}
2\pi ft_*-2\Phi(t_*)=2\pi ft_c-2\Phi(t_c)+\int^{\pi f}_\infty d\omega\frac{2\pi f-2\omega}{\dot\omega}.
\end{equation}
Therefore the phase $\Psi_+$ takes the following form:
\begin{equation}\label{psi}
\Psi_+=2\pi f(R+t_c)-2\Phi(t_c)-\pi/4+\int^{\pi f}_\infty d\omega\frac{2\pi f-2\omega}{\dot\omega}.
\end{equation}

To evaluate the integral in Eq.~\eqref{psi}, we must use the expression for the orbital frequency evolution given in Eq.~\eqref{omegadt}. However, this expression significantly depends on the value of the scalar field mass $m_\phi$. 

First, we estimate critical scalar field mass $\tilde m_\phi$ corresponding to $\tilde m_\phi r=1$. To perform this procedure we first estimate the parameter $r$ which is the relative distance between the binary system. Using the quasicircular equation of motion $v^2 =\mathcal{G}m/r$ with $v = r\omega$ and $\omega = 2\pi f$, and taking into accaunt that $r_g=\mathcal{G}m/2$, we obtain:
\begin{equation}
\tilde m_\phi=\frac{1}{r}=\biggl(\frac{8\pi^2f^2}{r_g}\biggr)^{1/3}\simeq10^{-12}eV\biggl(\frac{f}{50 Hz}\biggr)^{2/3}\biggl(\frac{r_g}{10^4 m}\biggr)^{-1/3}.
\end{equation}
Further we consider two cases $m_\phi\ll\tilde m_\phi$ and $m_\phi\gg\tilde m_\phi$. 

\subsection{$m_\phi\ll\tilde m_\phi$}

It was previously shown that in the case of a light scalar field its mass lies within the range $m_\phi<6.2\times10^{-18}$ eV \cite{Dyadina2018}. Therefore, we can safely assume that in this case $m_\phi r\ll1$ and $e^{-m_\phi r}=0$.

Generally speaking, scalar gravitational radiation exists (i.e., the scalar mode is excited) only if the frequency (energy) of the scalar mode exceeds its mass. In HMPG the Compton wavelength, $m^{-1}_\phi$ is typically of cosmological scale. For instance, if $m^{-1}_\phi\sim1$ Mpc, then $m_\phi\sim10^{-14}$ Hz. On the other hand, the characteristic orbital frequency of compact binary systems is $\omega\simeq100$ Hz. It follows that $m_\phi\ll\omega$ for such systems. 
 
 Under these assumptions, we can simplify Eq.~\eqref{omegadt} by taking into accaunt that $m_\phi r\ll1$ and $m_\phi\ll\omega$, and thus obtain the following expression for $\dot \omega$:
\begin{eqnarray}\label{omegadt1}
&\dot \omega=\frac{96(k^2M_c)^{5/3}\omega^{11/3}}{5(8\pi)^{5/3}(\phi_0+1)^{5/3}}\biggl[1-\frac{2}{3}\delta-\frac{5\phi_0 }{288(\mathcal{G}m\omega)^{2/3}}A_d^2-\frac{5\phi_0\mu}{144\mathcal{G}m}A_d\bar{A}_{d}\nonumber\\
&-\frac{\phi_0}{18}A_q^2 +\frac{\phi_0}{576}A_dA_o \biggr],
\end{eqnarray}
where we use $\mathcal{G}=\frac{k^2}{8\pi(1+\phi_0)}\biggl(1-\delta\biggr)$, $\delta=\frac{\phi_0}{3}\bigl(1-2s_a(\phi_0+1)\bigr)\bigl(1-2s_b(\phi_0+1)\bigr)$ within the framework of the imposed approximations. We use notation $\delta$ to avoid mixing contributions from scalar and tensor quadrupole radiation. Then,
\begin{eqnarray}
&1/\dot\omega\simeq (5/96)(k^2M_c)^{-5/3}(8\pi)^{5/3}\omega^{-11/3}(\phi_0+1)^{5/3}\biggl[1+\frac{2}{3}\delta\nonumber\\
&+\frac{5\phi_0 }{288(\mathcal{G}m\omega)^{2/3}}A_d^2+\frac{5\phi_0\mu}{144\mathcal{G}m}A_d\bar{A}_{d}+\frac{\phi_0}{18}A_q^2 -\frac{\phi_0}{576}A_dA_o \biggr]
\end{eqnarray}

 It follows that phase terms after integration take the form:
\begin{eqnarray}\label{psi1}
\Psi_+&=&\Psi_\times-\frac{\pi}{2}=2\pi f(R+t_c)-2\Phi(t_c)-\pi/4\nonumber\\
&+&\frac{3}{128}\biggl(\frac{k^2M_c\pi f}{8\pi(1+\phi_0)}\biggr)^{-5/3}\biggl[1+\frac{2}{3}\delta+\frac{5\phi_0 }{504(\mathcal{G}m\pi f)^{2/3}}A_d^2\nonumber\\
&+&\frac{5\phi_0\mu}{144\mathcal{G}m}A_d\bar{A}_{d}+\frac{\phi_0}{18}A_q^2 -\frac{\phi_0}{576}A_dA_o \biggr].
\end{eqnarray}
Here, we have ignored corrections of order  $\phi_0^2$ and higher.  As result we obtain
\begin{eqnarray}\label{h4}
&\tilde h_+(f)=-\frac{(k^2M_c)^{5/6}(1+\cos^2i)}{R[8\pi(\phi_0+1)]^{5/6}}\sqrt{\frac{5\pi}{96}}(\pi f)^{-7/6}\biggl[1-\frac{1}{3}\delta\\
&+\frac{5\phi_0 }{576(\mathcal{G}m\pi f)^{2/3}}A_d^2+\frac{5\phi_0\mu}{288\mathcal{G}m}A_d\bar{A}_{d}+\frac{\phi_0}{36}A_q^2 -\frac{\phi_0}{1152}A_dA_o\biggr]e^{i\Psi_+},\nonumber
\end{eqnarray}
\begin{eqnarray}\label{h5}
&\tilde h_\times(f)=-2\frac{(k^2M_c)^{5/6}\cos i}{R[8\pi(\phi_0+1)]^{5/6}}\sqrt{\frac{5\pi}{96}}(\pi f)^{-7/6}\biggl[1-\frac{1}{3}\delta\\
&+\frac{5\phi_0 }{576(\mathcal{G}m\pi f)^{2/3}}A_d^2+\frac{5\phi_0\mu}{288\mathcal{G}m}A_d\bar{A}_{d}+\frac{\phi_0}{36}A_q^2 -\frac{\phi_0}{1152}A_dA_o\biggr]e^{i\Psi_\times}.\nonumber
\end{eqnarray}

Thus, we conclude that all additional contributions to the tensor amplitudes, originating from scalar radiation (including dipole, quadrupole, and dipole-octupole terms), as well as from corrections to the gravitational constant, are suppressed due to the small background value of the scalar field. The background scalar field value is estimated to be \( \phi_0 \sim 10^{-5} \) according to the most stringent constraints \cite{Leanizbarrutia2017, Dyadina2019}. It has also been shown that the amplitude of the scalar breathing mode is of the same order. Consequently, any deviations arising when describing the gravitational-wave background within the framework of HMPG, compared to GR, are also expected to be of order \( 10^{-5} \). At present, detecting such small differences between these theories remains extremely challenging.

\subsection{$m_\phi\gg\tilde m_\phi$}

In this case all Heaviside functions in the Eq.~\eqref{omegadt} are equal to zero and scalar radiation vanish. Also we assume that $m_\phi r\gg1$. Then $\dot\omega=\frac{12k^2\mu(\mathcal{G}m)^{2/3}\omega^{11/3}}{5\pi (\phi_0+1
) }$, where $\mathcal{G}=\frac{k^2}{8\pi(1+\phi_0)}$. Thus for heavy scalar field we have:
\begin{eqnarray}\label{psi1''}
\Psi_+=&\Psi_\times-\frac{\pi}{2}=2\pi f(R+t_c)-2\Phi(t_c)-\pi/4\nonumber\\
&+\frac{3}{128}\biggl(\frac{k^2M_c\pi f}{8\pi(1+\phi_0)}\biggr)^{-5/3}.
\end{eqnarray}
\begin{eqnarray}\label{h4''}
\tilde h_+(f)=-\frac{(k^2M_c)^{5/6}(1+\cos^2i)}{R[8\pi(\phi_0+1)]^{5/6}}\sqrt{\frac{5\pi}{96}}(\pi f)^{-7/6}e^{i\Psi_+},
\end{eqnarray}
\begin{eqnarray}\label{h5''}
\tilde h_\times(f)=-2\frac{(k^2M_c)^{5/6}\cos i}{R[8\pi(\phi_0+1)]^{5/6}}\sqrt{\frac{5\pi}{96}}(\pi f)^{-7/6}e^{i\Psi_\times}.
\end{eqnarray}

Thus, in the case of a heavy scalar field, scalar radiation is absent, and the only deviation from the predictions of GR arises from the effective redefinition of the gravitational constant due to the background value of the scalar field. Since independent observational data also constrain $|\phi_0|<10^{-4}$ in this regime, detecting any deviation of HMPG from GR remains an extremely challenging task.

\section{Parametrized post-Einstein parameters}\label{ppe}

The parametrized post-Einsteinian (ppE) framework was proposed by Yunes and Pretorius \cite{Yunes_2009} to describe GWs emitted by binary system on quasi-circular orbits in metric theories of gravity. Within ppE formalism, all deviations from GR in gravitational waveforms can be expressed through a set of four post-Einstein parameters
$(\alpha_{ppe},\beta_{ppe},a,b)$. It is worth to noting that the original ppE framework includes only the two tensor polarizations $h_+$ and $h_\times$. At present, a more general extension of the ppE formalism has been developed, which introduces a broader set of parameters and allows for a wider range of deviations from GR in both the amplitude and phase of GWs.  A more general the gravitational waveform $\tilde h(f)$ can be expressed as 
\begin{equation}
\tilde h(f)=\tilde h_{GR}(f)\bigl(1+\Sigma_j\alpha_{j}(GM_c\pi f)^{a_j/3}\bigr)e^{i\Sigma_j\beta_{j}(GM_c\pi f)^{b_j/3}},
\end{equation}
where $\tilde h_{GR}(f)$  represents the Fourier waveform according to GR, while  $(\alpha_{j},\beta_{j}, a_j, b_j)$ denote the set of ppE parameters characterizing deviations due to modified gravity effects.

Further, we focus on the leading-order correction to the gravitational wave phase and amplitude induced by scalar dipole radiation. In the light mass regime $m_\phi\ll\omega$ according to Eqs.~\eqref{psi1}, \eqref{h4}, \eqref{h5}, we obtain the following set of ppE parameters in HMPG:
\begin{eqnarray}
&\alpha_1=\frac{5}{576}\phi_0A_d^2\eta^{2/5},\ \ \ \ \beta_1=\frac{5}{21504}\phi_0A_d^2\eta^{2/5},\nonumber\\
 &a_1=-2,\ \ \ \ \ b_1=-7,
\end{eqnarray}
where $\eta=\mu/m$.

In the work \cite{Chamberlain_2017} authors investigated the possible observational constraints on ppE parameter $\beta$  using future ground-based gravitational-wave detectors such as the LIGO-class expansions A+, Voyager, Cosmic Explorer, and the Einstein Telescope, as well as various configurations the space-based detector LISA. Their analysis focused on GWs from mixed binary systems consisting of a neutron star with $m_{NS}=1.4 M_\odot$ and a black hole with $m_{BH}=5M_\odot$ located at a distance of 150 Mpc. As a result authors derived the constraints on $|\beta_{ppe}|$. We are interested only in $\beta_{ppe}$, which corresponds to $b_{ppe}=-7$. In this case we have $|\beta_{ppe}|<2.88\times10^{-8}$ \cite{Chamberlain_2017, Liu_2020}. Using this bound, we can constrain $\phi_0$ as:
	\begin{equation}
	|\phi_0|<7\times10^{-4}.
	\end{equation}
	
	The resulting constraint is less stringent than those derived earlier from Solar System observations \cite{Leanizbarrutia2017, Dyadina2019} and from our analysis based on updated parameters of the double pulsar system.

\section{Conclusion}\label{concl}

In this work, we studied gravitational radiation from quasi-circular binary systems with compact objects within the hybrid metric-Palatini gravity. The main goal of the article was to calculate the gravitational waveforms emitted during the inspiral phase of such systems in the lowest post-Newtonian approximation. HMPG predicts the existence of two tensor  ($h_+$ and $h_\times$) and two scalar ($h_b$ and $h_L$) GW polarizations \cite{kausar2018, Dyadina2022}. We derived analytical expressions for the amplitudes of all these modes. Besides, we found that scalar radiation exists only when the scalar field is light  ($m_\phi\ll\omega$). In this case, we calculated ratio between the scalar breathing mode $h_b$ and the tensor $h_+$ mode, determining that $h_b$ is $10^5$ times weaker than $h_+$. This suppression of the scalar breathing mode is due to the small background value of the scalar field, which has been constrained previously within the Solar System \cite{Leanizbarrutia2017, Dyadina2019}. Moreover, the smaller the value of $\phi_0$ we take, the greater the gap between the magnitudes of the amplitudes of scalar and tensor radiation. Additionally, we evaluated the relationship between scalar modes and discovered that the scalar longitudinal mode is $10^{12}$ times smaller than the scalar breathing mode. A linear relationship was identified between these modes, where the coupling coefficient is $m_\phi^2/4\omega^2$. Thus, only at frequencies comparable to the mass of the scalar field $\omega\gtrsim m_\phi$, does the longitudinal component of scalar radiation become comparable to the scalar breathing mode. 

Further, using the SPA method, we computed the Fourier transforms of the two tensor GW polarizations $\tilde h_+(f)$ and $\tilde h_\times(f)$. We also derived analytical expressions for the phases of these polarizations $\Psi_+$ and $\Psi_\times$ in HMPG. Our calculations relied on earlier results from \cite{Dyadina2018}, particularly using the expression for the orbital period change $\dot{P}_b$ in compact binary systems and consequently the orbital frequency change $\dot{\omega}$.  Moreover, in this work, we take into account the contribution of sensitivities to the $\dot{P}_b$ expression and their significant impact on the description of strongly gravitating objects. Separately, we considered the case of a light scalar field and found that the correction due to the presence of scalar radiation is suppressed due to the small background value of the scalar field. This circumstance complicates the detection of deviations from GR, especially given the current observational accuracy. To verify the last statement, we identified the ppE parameters within the framework of HMPG. Considering the potential observations of the GWs emitted by a black hole-neutron star binary by future ground-based GW detectors \cite{Chamberlain_2017, Liu_2020}, we obtained the constraints on the parameter $\phi_0$. As a result, we determined that these constraints are  less stringent compared to those obtained from observations within the Solar System.

In addition, it should be noted that some of the results obtained within the framework of hybrid metric-Palatini gravity can be formally derived from the corresponding expressions of the massive Brans-Dicke theory presented in Ref.~\cite{Liu_2020} via conformal transformations, in particular, the expressions for the effective gravitational constant $G_{eff}$ and the amplitudes of gravitational waves $h_+, h_\times, h_{b1}, h_{b2}, h_{L1}, h_{L2}$. However, when performing such transformations, it is essential to take into account the fundamentally different definitions of sensitivities of compact objects in these theories. In Brans-Dicke theory, sensitivities are traditionally defined through the logarithmic derivative of the scalar field, which leads to large scalar contributions due to the small values of the field in the denominator in HMPG framework. Therefore, in HMPG, it is natural to use an alternative definition of sensitivities directly in terms of the scalar field, without taking its logarithm. Taking these subtleties into account leads to a mathematical agreement between the expressions; however, the physical equivalence of HMPG and massive Brans-Dicke theories remains an open question and requires a separate detailed investigation. A similar situation is well known in Palatini $f(R)$-gravity, where the mathematical equivalence to Brans-Dicke theory does not result in physical identity due to differences in assumptions regarding free fall and the interpretation of observable quantities \cite{Fatibene:2013oka}. The agreement between the formulas obtained in our work and in Ref.~\cite{Liu_2020} provides additional confirmation of the internal consistency of the proposed approach and highlights the need for further studies of the physical differences between the mathematically equivalent but potentially physically distinct formulations of the theory.

Another direction of this study was to establish constraints on the background value of the scalar field using updated observational data from the PSR J0737-3039 system, the only known double pulsar.  Previous research, as outlined in  \cite{Dyadina2018}, had already utilized observational data on orbital period changes to impose restrictions on HMPG.  Recently published updated data have significantly enhanced accuracy of these measurements \cite{Kramer_2021}. In our study, we relied on the methods outlined in paper \cite{Dyadina2018}, but additionally accounted for the contribution of sensitivities when deriving the expression for the orbital period change, which was not considered previously. Furthermore, using updated values of  observational parameters, we obtained improved constraints on the background value of the scalar field. This limitation is comparable in order of magnitude to the best currently existing constraints, which were derived from Cassini data.  While these constraints are slightly less stringent than those obtained from the Solar System \cite{Leanizbarrutia2017, Dyadina2019}, continued observations of the system PSR J0737-3039 will further refine the limits on $\phi_0$. Additionally, performing a full post-Keplerian test may improve existing constraints. Moreover, we investigated the case of heavy scalar field and obtained the most stringent constraints currently available for this case.

As part of this study, we have demonstrated that the presence of a light scalar field in HMPG does not lead to significant deviations from GR in the description of GWs. The gravitational waveforms within the framework of HMPG remain entirely consistent with both the predictions of GR and current observations of gravitational-wave radiation, as well as with the expected outcomes from future detectors. In this context, any additional effects of HMPG, intended to explain the accelerated expansion of the Universe, are suppressed due to the small magnitude of the scalar field background value in local astrophysical systems. This work represents the first step toward the study of gravitational waveforms in HMPG, which will make it possible to further constrain the theory and gain deeper insights into the fundamental nature of gravity.
 
\begin{acknowledgements} 
 The author thanks N. A. Avdeev, G.A. Alekseev, T. Lui, E. Barausse for useful discussions.  The work was supported by the Foundation for the Advancement of Theoretical Physics and Mathematics “BASIS”.
 \end{acknowledgements}
 \section*{Declarations}

\textbf{Conflict of interest} The authors declare that they have no known competing financial interests or personal relationships that could have appeared to influence the work reported in this paper.

\textbf{Data Availability Statement} The data that support the findings of this study are available in the referenced publications.

\textbf{Code Availability Statement} This manuscript has no associated
code/software.

\bibliographystyle{spphys}
\bibliography{gravitational_waveforms.bib} 

\begin{thebibliography}{10}
\providecommand{\url}[1]{{#1}}
\providecommand{\urlprefix}{URL }
\expandafter\ifx\csname urlstyle\endcsname\relax
  \providecommand{\doi}[1]{DOI \discretionary{}{}{}#1}\else
  \providecommand{\doi}{DOI \discretionary{}{}{}\begingroup
  \urlstyle{rm}\Url}\fi

\bibitem{Abbott_2016}
B.~Abbott, et~al., Physical Review Letters \textbf{116}(13).
\newblock \doi{10.1103/physrevlett.116.131103}.
\newblock \urlprefix\url{http://dx.doi.org/10.1103/PhysRevLett.116.131103}

\bibitem{Abbott_2017}
B.~Abbott, et~al., Physical Review Letters \textbf{119}(16) (2017).
\newblock \doi{10.1103/physrevlett.119.161101}.
\newblock \urlprefix\url{http://dx.doi.org/10.1103/PhysRevLett.119.161101}

\bibitem{Perlmutter1999}
S.~Perlmutter, et~al., Astrophys.J. \textbf{517}(2), 565 (1999).
\newblock \doi{10.1086/307221}.
\newblock \urlprefix\url{https://ui.adsabs.harvard.edu/abs/1999ApJ...517..565P}

\bibitem{Riess1998}
A.G. Riess, et~al., Astron.J. \textbf{116}(3), 1009 (1998).
\newblock \doi{10.1086/300499}.
\newblock \urlprefix\url{https://ui.adsabs.harvard.edu/abs/1998AJ....116.1009R}

\bibitem{zwicky}
F.~{Zwicky}, Helvetica Physica Acta \textbf{6}, 110 (1933)

\bibitem{oort}
J.H. {Oort}, Bulletin of the Astronomical Institutes of the Netherlands
  \textbf{6}, 249 (1932)

\bibitem{Starobinsky1980}
A.~Starobinsky, Physics Letters B \textbf{91}(1), 99 (1980).
\newblock \doi{10.1016/0370-2693(80)90670-x}

\bibitem{Guth1981}
A.H. Guth, Physical Review D \textbf{23}(2), 347 (1981).
\newblock \doi{10.1103/physrevd.23.347}

\bibitem{Linde1982}
A.~Linde, Physics Letters B \textbf{108}(6), 389 (1982).
\newblock \doi{10.1016/0370-2693(82)91219-9}

\bibitem{Burgess_2004}
C.P. Burgess, Living Reviews in Relativity \textbf{7}(1) (2004).
\newblock \doi{10.12942/lrr-2004-5}.
\newblock \urlprefix\url{http://dx.doi.org/10.12942/lrr-2004-5}

\bibitem{Latosh_2020}
B.~Latosh, The European Physical Journal C \textbf{80}(9) (2020).
\newblock \doi{10.1140/epjc/s10052-020-8371-2}.
\newblock \urlprefix\url{http://dx.doi.org/10.1140/epjc/s10052-020-8371-2}

\bibitem{Will_2014}
C.M. Will, Living Reviews in Relativity \textbf{17}(1) (2014).
\newblock \doi{10.12942/lrr-2014-4}.
\newblock \urlprefix\url{http://dx.doi.org/10.12942/lrr-2014-4}

\bibitem{kopeikin}
S.~{Kopeikin}, M.~{Efroimsky}, G.~{Kaplan}, \emph{{Relativistic Celestial
  Mechanics of the Solar System}} (John Wiley \& Sons, Ltd, 2011).
\newblock \doi{10.1002/9783527634569}

\bibitem{will1993theory}
C.M. Will, \emph{Theory and Experiment in Gravitational Physics} (Cambridge
  University Press, 1993)

\bibitem{Bergmann1968}
P.G. Bergmann, International Journal of Theoretical Physics \textbf{1}(1), 25
  (1968).
\newblock \doi{10.1007/bf00668828}

\bibitem{Felice2010}
A.D. Felice, S.~Tsujikawa, Living Reviews in Relativity \textbf{13}(1) (2010).
\newblock \doi{10.12942/lrr-2010-3}

\bibitem{Nojiri2017}
S.~Nojiri, S.~Odintsov, V.~Oikonomou, Physics Reports \textbf{692}, 1 (2017).
\newblock \doi{10.1016/j.physrep.2017.06.001}

\bibitem{Nojiri2011}
S.~Nojiri, S.D. Odintsov, Physics Reports \textbf{505}(2-4), 59 (2011).
\newblock \doi{10.1016/j.physrep.2011.04.001}

\bibitem{Chiba2003}
T.~Chiba, Physics Letters B \textbf{575}(1-2), 1 (2003).
\newblock \doi{10.1016/j.physletb.2003.09.033}

\bibitem{Olmo2005}
G.J. Olmo, Physical Review Letters \textbf{95}(26), 261102 (2005).
\newblock \doi{10.1103/physrevlett.95.261102}

\bibitem{Olmo2007}
G.J. Olmo, Physical Review D \textbf{75}(2), 023511 (2007).
\newblock \doi{10.1103/physrevd.75.023511}

\bibitem{odintsov1}
S.~Nojiri, S.D. Odintsov, Phys. Rev. D \textbf{77}, 026007 (2008).
\newblock \doi{10.1103/PhysRevD.77.026007}.
\newblock \urlprefix\url{https://link.aps.org/doi/10.1103/PhysRevD.77.026007}

\bibitem{odintsov2}
G.~Cognola, E.~Elizalde, S.~Nojiri, S.D. Odintsov, L.~Sebastiani, S.~Zerbini,
  Phys. Rev. D \textbf{77}, 046009 (2008).
\newblock \doi{10.1103/PhysRevD.77.046009}.
\newblock \urlprefix\url{https://link.aps.org/doi/10.1103/PhysRevD.77.046009}

\bibitem{odintsov3}
S.D. Odintsov, D.S.C. G{\'{o}}mez, G.S. Sharov, The European Physical Journal C
  \textbf{77}(12) (2017).
\newblock \doi{10.1140/epjc/s10052-017-5419-z}.
\newblock \urlprefix\url{https://doi.org/10.1140%2Fepjc%2Fs10052-017-5419-z}

\bibitem{koivisto2006}
T.~Koivisto, H.~Kurki-Suonio, Classical and Quantum Gravity \textbf{23}(7),
  2355 (2006).
\newblock \doi{10.1088/0264-9381/23/7/009}

\bibitem{koivisto206}
T.~Koivisto, Phys. Rev. D \textbf{73}, 083517 (2006).
\newblock \doi{10.1103/PhysRevD.73.083517}.
\newblock \urlprefix\url{https://link.aps.org/doi/10.1103/PhysRevD.73.083517}

\bibitem{Harko2012}
T.~Harko, T.S. Koivisto, F.S.N. Lobo, G.J. Olmo, Physical Review D
  \textbf{85}(8), 084016 (2012).
\newblock \doi{10.1103/physrevd.85.084016}

\bibitem{Capozziello2015}
S.~Capozziello, T.~Harko, T.~Koivisto, F.~Lobo, G.~Olmo, Universe
  \textbf{1}(2), 199 (2015).
\newblock \doi{10.3390/universe1020199}

\bibitem{Harko2020}
T.~Harko, F.S.N. Lobo, International Journal of Modern Physics D
  \textbf{29}(13), 2030008 (2020).
\newblock \doi{10.1142/s0218271820300086}.
\newblock \urlprefix\url{https://doi.org/10.1142%2Fs0218271820300086}

\bibitem{Boehmer2013}
C.G. Böhmer, F.S.N. Lobo, N.~Tamanini, Physical Review D \textbf{88}(10),
  104019 (2013).
\newblock \doi{10.1103/physrevd.88.104019}

\bibitem{Lima2016}
N.A. Lima, V.~Smer-Barreto, L.~Lombriser, Physical Review D \textbf{94}(8),
  083507 (2016).
\newblock \doi{10.1103/physrevd.94.083507}

\bibitem{Capozziello2013}
S.~Capozziello, T.~Harko, T.S. Koivisto, F.S. Lobo, G.J. Olmo, Journal of
  Cosmology and Astroparticle Physics \textbf{2013}(04), 011 (2013).
\newblock \doi{10.1088/1475-7516/2013/04/011}

\bibitem{Capozziello2013a}
S.~Capozziello, T.~Harko, T.S. Koivisto, F.S. Lobo, G.J. Olmo, Astroparticle
  Physics \textbf{50-52}, 65 (2013).
\newblock \doi{10.1016/j.astropartphys.2013.09.005}

\bibitem{Capozziello13}
S.~Capozziello, T.~Harko, T.S. Koivisto, F.S. Lobo, G.J. Olmo, Journal of
  Cosmology and Astroparticle Physics \textbf{2013}(07), 024 (2013).
\newblock \doi{10.1088/1475-7516/2013/07/024}.
\newblock \urlprefix\url{https://doi.org/10.1088%2F1475-7516%2F2013%2F07%2F024}

\bibitem{Leanizbarrutia2017}
I.~Leanizbarrutia, F.S. Lobo, D.~S{\'{a}}ez-G{\'{o}}mez, Physical Review D
  \textbf{95}(8), 084046 (2017).
\newblock \doi{10.1103/physrevd.95.084046}.
\newblock \urlprefix\url{https://ui.adsabs.harvard.edu/abs/2017PhRvD..95h4046L}

\bibitem{Dyadina2019}
P.I. Dyadina, S.P. Labazova, S.O. Alexeyev, Journal of Experimental and
  Theoretical Physics \textbf{129}(5), 838 (2019).
\newblock \doi{10.1134/s1063776119110025}

\bibitem{Dyadina_2022}
P.~Dyadina, S.~Labazova, Journal of Cosmology and Astroparticle Physics
  \textbf{2022}(01), 029 (2022).
\newblock \doi{10.1088/1475-7516/2022/01/029}.
\newblock \urlprefix\url{https://doi.org/10.1088%2F1475-7516%2F2022%2F01%2F029}

\bibitem{Dyadina2018}
P.I. Dyadina, N.A. Avdeev, S.O. Alexeyev, Monthly Notices of the Royal
  Astronomical Society \textbf{483}(1), 947 (2018).
\newblock \doi{10.1093/mnras/sty3094}

\bibitem{Avdeev2020}
N.A. Avdeev, P.I. Dyadina, S.P. Labazova, Journal of Experimental and
  Theoretical Physics \textbf{131}(4), 537 (2020).
\newblock \doi{10.1134/s1063776120100039}

\bibitem{Danil2017}
B.~Danil{\u{a} }, T.~Harko, F.S. Lobo, M.~Mak, Physical Review D \textbf{95}(4)
  (2017).
\newblock \doi{10.1103/physrevd.95.044031}.
\newblock \urlprefix\url{https://doi.org/10.1103%2Fphysrevd.95.044031}

\bibitem{Danila2019}
B.~Dǎnilǎ, T.~Harko, F.S. Lobo, M.K. Mak, Physical Review D \textbf{99}(6),
  064028 (2019).
\newblock \doi{10.1103/physrevd.99.064028}

\bibitem{Bronnikov2020}
K.A. Bronnikov, S.V. Bolokhov, M.V. Skvortsova, Gravitation and Cosmology
  \textbf{26}(3), 212 (2020).
\newblock \doi{10.1134/s0202289320030044}

\bibitem{Dyadina_2024}
P.~Dyadina, N.~Avdeev, The European Physical Journal C \textbf{84}(1) (2024).
\newblock \doi{10.1140/epjc/s10052-024-12465-7}.
\newblock \urlprefix\url{http://dx.doi.org/10.1140/epjc/s10052-024-12465-7}

\bibitem{Rosa2017}
J.L. Rosa, S.~Carloni, J.P.S. Lemos, F.S.N. Lobo, Physical Review D
  \textbf{95}(12) (2017).
\newblock \doi{10.1103/physrevd.95.124035}.
\newblock \urlprefix\url{https://doi.org/10.1103%2Fphysrevd.95.124035}

\bibitem{Rosa20}
J.a.L. Rosa, J.P.S. Lemos, F.S.N. Lobo, Phys. Rev. D \textbf{101}, 044055
  (2020).
\newblock \doi{10.1103/PhysRevD.101.044055}.
\newblock \urlprefix\url{https://link.aps.org/doi/10.1103/PhysRevD.101.044055}

\bibitem{Rosa2021}
J.L. Rosa, F.S.N. Lobo, G.J. Olmo, Physical Review D \textbf{104}(12) (2021).
\newblock \doi{10.1103/physrevd.104.124030}.
\newblock \urlprefix\url{https://doi.org/10.1103%2Fphysrevd.104.124030}

\bibitem{Liu_2020}
T.~Liu, W.~Zhao, Y.~Wang, Physical Review D \textbf{102}(12) (2020).
\newblock \doi{10.1103/physrevd.102.124035}.
\newblock \urlprefix\url{http://dx.doi.org/10.1103/PhysRevD.102.124035}

\bibitem{Higashino_2023}
Y.~Higashino, S.~Tsujikawa, Physical Review D \textbf{107}(4) (2023).
\newblock \doi{10.1103/physrevd.107.044003}.
\newblock \urlprefix\url{http://dx.doi.org/10.1103/PhysRevD.107.044003}

\bibitem{Liu_2018}
T.~Liu, X.~Zhang, W.~Zhao, K.~Lin, C.~Zhang, S.~Zhang, X.~Zhao, T.~Zhu,
  A.~Wang, Physical Review D \textbf{98}(8) (2018).
\newblock \doi{10.1103/physrevd.98.083023}.
\newblock \urlprefix\url{http://dx.doi.org/10.1103/PhysRevD.98.083023}

\bibitem{kausar2018}
H.R. Kausar, Astrophys. Space Sci. \textbf{363}(11), 238 (2018).
\newblock \doi{10.1007/s10509-018-3458-z}

\bibitem{Dyadina2022}
P.I. Dyadina, Journal of Experimental and Theoretical Physics \textbf{135}(3),
  333 (2022).
\newblock \doi{10.1134/s106377612208009x}.
\newblock \urlprefix\url{https://doi.org/10.1134%2Fs106377612208009x}

\bibitem{Kramer_2021}
M.~Kramer, et~al., Physical Review X \textbf{11}(4) (2021).
\newblock \doi{10.1103/physrevx.11.041050}.
\newblock \urlprefix\url{http://dx.doi.org/10.1103/PhysRevX.11.041050}

\bibitem{Bronnikov_2019}
K.A. Bronnikov, Gravitation and Cosmology \textbf{25}(4), 331–341 (2019).
\newblock \doi{10.1134/s0202289319040030}.
\newblock \urlprefix\url{http://dx.doi.org/10.1134/S0202289319040030}

\bibitem{PhysRevD.8.3308}
D.M. Eardley, D.L. Lee, A.P. Lightman, Phys. Rev. D \textbf{8}, 3308 (1973).
\newblock \doi{10.1103/PhysRevD.8.3308}.
\newblock \urlprefix\url{https://link.aps.org/doi/10.1103/PhysRevD.8.3308}

\bibitem{Barausse_2015}
E.~Barausse, K.~Yagi, Physical Review Letters \textbf{115}(21) (2015).
\newblock \doi{10.1103/physrevlett.115.211105}.
\newblock \urlprefix\url{http://dx.doi.org/10.1103/PhysRevLett.115.211105}

\bibitem{Cassini2003}
B.~Bertotti, L.~Iess, P.~Tortora, Nature \textbf{425}, 374 (2003).
\newblock \doi{10.1038/nature01997}

\bibitem{Alsing_2012}
J.~Alsing, E.~Berti, C.M. Will, H.~Zaglauer, Physical Review D \textbf{85}(6)
  (2012).
\newblock \doi{10.1103/physrevd.85.064041}.
\newblock \urlprefix\url{http://dx.doi.org/10.1103/PhysRevD.85.064041}

\bibitem{maggiore2007gravitational}
M.~Maggiore, \emph{Gravitational Waves. Vol. 1: Theory and Experiments}.
\newblock Oxford Master Series in Physics (Oxford University Press, 2007).
\newblock \doi{https://doi.org/10.1093/acprof:oso/9780198570745.001.0001}

\bibitem{Maggiore_2000}
M.~Maggiore, A.~Nicolis, Physical Review D \textbf{62}(2) (2000).
\newblock \doi{10.1103/physrevd.62.024004}.
\newblock \urlprefix\url{http://dx.doi.org/10.1103/PhysRevD.62.024004}

\bibitem{Yunes_2009}
N.~Yunes, F.~Pretorius, Physical Review D \textbf{80}(12) (2009).
\newblock \doi{10.1103/physrevd.80.122003}.
\newblock \urlprefix\url{http://dx.doi.org/10.1103/PhysRevD.80.122003}

\bibitem{Chamberlain_2017}
K.~Chamberlain, N.~Yunes, Physical Review D \textbf{96}(8) (2017).
\newblock \doi{10.1103/physrevd.96.084039}.
\newblock \urlprefix\url{http://dx.doi.org/10.1103/PhysRevD.96.084039}

\bibitem{Fatibene:2013oka}
L.~Fatibene, M.~Francaviglia, Int. J. Geom. Meth. Mod. Phys. \textbf{11},
  1450008 (2014).
\newblock \doi{10.1142/S021988781450008X}

\end{thebibliography}

\end{document}